\documentclass[twocolumn,astrosymb]{aastex701}

\usepackage{graphicx} 
\usepackage{xspace}
\usepackage{amsmath}
\usepackage{placeins}
\usepackage{hyperref}
\usepackage{comment}
\usepackage{upgreek}
\usepackage{color}
\usepackage{natbib}

\newcommand{\um}{\ensuremath{\upmu\mathrm{m}}\xspace}
\newcommand{\csixty}{C$_{60}$\xspace}
\newcommand{\cseventy}{C$_{70}$\xspace}
\newcommand{\csixtyplus}{C$_{60}^{+}$\xspace}

\newcommand{\ArII}{\textrm{Ar~{\textsc{ii}}}}
\newcommand{\HII}{\textrm{H~{\textsc{ii}}}\xspace}
\newcommand{\KrIII}{\textrm{Kr~{\textsc{iii}}}\xspace}

\definecolor{cbpurple}{rgb}{0.47, 0.37, 0.94}

\newcommand{\overviewpaper}{Paper~I\xspace}
\newcommand{\morganpaper}{Paper~II\xspace}
\newcommand{\citePAHPaper}{Van~Schuylenbergh et al. (in review)\xspace}
\newcommand{\citepPAHPaper}{(Van~Schuylenbergh et al. in review)\xspace}

\begin{document}

\title{Studying the Fullerene Nest: JWST Observations of the Planetary Nebula Tc 1. \\III. Tracing Fullerene-Linked Dust Variation}

\correspondingauthor{Jan Cami}
\email{jcami@uwo.ca}

\author[0009-0004-9270-0019]{Simon Van Schuylenbergh}
\affiliation{Department of Physics and Astronomy, The University of Western Ontario, London, Canada, N6A 3K7}
\affiliation{Institute for Earth and Space Exploration, The University of Western Ontario, London, Canada, N6A 3K7}
\email{svanschu@uwo.ca}

\author[0000-0002-2666-9234]{Jan Cami}
\affiliation{Department of Physics and Astronomy, The University of Western Ontario, London, Canada, N6A 3K7}
\affiliation{Institute for Earth and Space Exploration, The University of Western Ontario, London, Canada, N6A 3K7}
\email{jcami@uwo.ca}

\author[0000-0002-2541-1602]{Els Peeters}
\affiliation{Department of Physics and Astronomy, The University of Western Ontario, London, Canada, N6A 3K7}
\affiliation{Institute for Earth and Space Exploration, The University of Western Ontario, London, Canada, N6A 3K7}
\email{epeeters@uwo.ca}

\author[0009-0007-3950-335X]{Morgan M. Giese}
\affiliation{Department of Physics and Astronomy, The University of Western Ontario, London, Canada, N6A 3K7}
\affiliation{Institute for Earth and Space Exploration, The University of Western Ontario, London, Canada, N6A 3K7}
\email{mgiese@uwo.ca}

\author[0009-0000-0191-6756]{Charmi Bhatt}
\affiliation{Department of Physics and Astronomy, The University of Western Ontario, London, Canada, N6A 3K7}
\affiliation{Institute for Earth and Space Exploration, The University of Western Ontario, London, Canada, N6A 3K7}
\email{cbhatt7@uwo.ca}

\author[0000-0002-5895-8268]{Dries Van De Putte}
\affiliation{Department of Physics and Astronomy, The University of Western Ontario, London, Canada, N6A 3K7}
\affiliation{Institute for Earth and Space Exploration, The University of Western Ontario, London, Canada, N6A 3K7}
\email{dvandepu@uwo.ca}

\author{Cedric Baerts}
\affiliation{Department of Physics and Astronomy, The University of Western Ontario, London, Canada, N6A 3K7}
\affiliation{Institute for Earth and Space Exploration, The University of Western Ontario, London, Canada, N6A 3K7}
\email{cbaerts@uwo.ca}

\author[0000-0002-3875-1171]{Michael J. Barlow}
\affiliation{Department of Physics and Astronomy, University College London, Gower Street, London WC1E 6BT, United Kingdom}
\email{mjb@star.ucl.ac.uk}

\author[0000-0002-8452-8675]{Jeronimo Bernard-Salas}
\affiliation{ACRI-ST, Centre d'Etudes et de Recherche de Grasse (CERGA), 10Av. Nicolas Copernic, 06130 Grasse, France}
\affiliation{INCLASS Common Laboratory, 10 Av. Nicolas Copernic, 06130 Grasse, France}
\email{jeronimo.bernard-salas@acri-st.fr}

\author[0000-0002-5431-4449]{Alessandra Candian}
\affiliation{Anton Pannekoek Institute, University of Amsterdam, Science Park 904, 1098XH Am-
sterdam, The Netherlands}
\email{a.candian2@uva.nl}

\author[0000-0003-0304-9814]{P. Bryan Changala}
\affiliation{JILA, University of Colorado Boulder and National Institute of Standards and Technology, Boulder, CO 80309, USA}
\affiliation{Department of Physics, University of Colorado Boulder, Boulder CO 80309}
\email{changala@jila.colorado.edu}

\author[0000-0002-7926-4492]{Nick L. J. Cox}
\affiliation{ACRI-ST, Centre d'Etudes et de Recherche de Grasse (CERGA), 10Av. Nicolas Copernic, 06130 Grasse, France}
\email{nljcox@gmail.com}

\author[0000-0002-4017-5572]{Harriet L. Dinerstein}
\affiliation{Department of Astronomy, University of Texas at Austin, Austin, TX 78712, USA }
\email{harriet@astro.as.utexas.edu}

\author[0000-0001-6035-3869]{Vincent J. Esposito}
\affiliation{Schmid College of Science and Technology, Chapman University, CA 92866, USA}
\email{vesposito@chapman.edu}

\author[0000-0002-1693-2721]{D. A. Garc\'ia-Hern\'andez}
\affiliation{Instituto de Astrof\'isica de Canarias, C/ Via L\'actea s/n, E-38205 La Laguna, Spain}
\affiliation{Departamento de Astrof\'isica, Universidad de La Laguna (ULL), E-38206 La Laguna, Spain}
\email{anibal.garcia.hernandez@iac.es}

\author[0000-0002-3938-4211]{Marco A. G\'omez-Mu\~noz}
\affiliation{Departament de F\'isica Qu\`antica i Astrof\'isica (FQA), Universitat de Barcelona (UB), \\c. Mart\'i i Franqu\`es, 1, 08028 Barcelona, Spain}
\affiliation{Institut de Ci\`encies del Cosmos (ICCUB), Universitat de Barcelona (UB), c. Mart\'i i Franqu\`es, 1, 08028 Barcelona, Spain}
\email{magm@iac.es}

\author[0000-0003-1689-9201]{Kay Justtanont}
\affiliation{Department of Physics and Astronomy, Chalmers University of Technology, 412 96 Gothenburg, Sweden }
\email{kay.justtanont@chalmers.se}

\author[0000-0002-2626-7155]{Kathleen E. Kraemer}
\affiliation{Institute for Scientific Research, Boston College, 140 Commonwealth Avenue, Chestnut Hill, MA 02467, USA}
\email{kathleen.kraemer@bc.edu}

\author[0000-0002-1335-5623]{Eric Lagadec}
\affiliation{Universit\'e C\^{o}te d’Azur, Observatoire de la C\^{o}te d’Azur, CNRS, Lagrange,
96 Bd de l’Observatoire, 06300 Nice, France }
\email{Eric.Lagadec@oca.eu}

\author[0000-0002-3011-686X]{Arturo Manchado}
\affiliation{Instituto de Astrof\'isica de Canarias, C/ V\'ia L\'actea s/n, E-38205 La Laguna, Spain}
\affiliation{Departamento de Astrof\'isica, Universidad de La Laguna, E-38206 La Laguna, Tenerife, Spain}
\affiliation{Consejo Superior de Investigaciones Científicas (CSIC), Spain}
\email{arturo.manchado@iac.es}

\author[0000-0002-6455-2491]{Ana Monreal Ibero}
\affiliation{Leiden Observatory, Leiden University, PO Box 9513, 2300 RA Leiden, The Netherlands}
\email{monreal@strw.leidenuniv.nl}

\author[0000-0002-6858-5063]{Raghvendra Sahai}
\affiliation{Jet Propulsion Laboratory, California Institute of Technology, Pasadena, CA, USA}
\email{raghvendra.sahai@jpl.nasa.gov}

\author[0000-0003-3771-4990]{Ameek Sidhu}
\affiliation{Department of Physics and Astronomy, The University of Western Ontario, London, Canada, N6A 3K7}
\affiliation{Institute for Earth and Space Exploration, The University of Western Ontario, London, Canada, N6A 3K7}
\email{asidhu92@uwo.ca}

\author[0000-0003-4520-1044]{G. C. Sloan}
\affiliation{Space Telescope Science Institute, 3700 San Martin Drive, Baltimore, MD 21218, USA}
\affiliation{Department of Physics and Astronomy, University of North Carolina, Chapel Hill, NC 27599-3255, USA }
\email{gcsloan@stsci.edu}

\author[0000-0002-9604-1434]{N. C. Sterling}
\affiliation{University of West Georgia, Carrollton, GA 30118, USA}
\email{nsterlin@westga.edu}

\author[0000-0003-0306-0028]{A.~G.~G.~M. Tielens}
\affiliation{Astronomy Department, University of Maryland, College Park, MD
20742, USA}
\email{tielens@strw.leidenuniv.nl}

\author[0000-0002-8008-910X]{Jeremy R. Walsh}
\affiliation{European Southern Observatory, Karl-Schwarzschild Strasse 2, 85748 Garching, Germany}
\email{jwalsh@eso.org}

\author[0000-0002-4000-4394]{Roger Wesson}
\affiliation{Department of Physics and Astronomy, University College London, Gower Street, London WC1E 6BT, United Kingdom}
\affiliation{Cardiff Hub for Astrophysics Research and Technology (CHART),
School of Physics and Astronomy, \\Cardiff University, The Parade, Cardiff CF24 3AA, UK}
\email{rw@nebulousresearch.org}

\author{Joshua Cole Whitman}
\affiliation{Department of Physics, University of West Georgia, Carrollton, GA 30118, USA}
\email{jwhitma3@my.westga.edu}

\author[0000-0002-3171-5469]{Albert Zijlstra}
\affiliation{Jodrell Bank Centre for Astrophysics, Department of Physics \&\ Astronomy, The University of Manchester, \\Oxford Road, Manchester M13 9PL, UK}
\email{a.zijlstra@manchester.ac.uk}

\begin{abstract}

We have observed the fullerene-rich planetary nebula Tc~1 with the MIRI/MRS on JWST. The mid-infrared spectra of this spatially resolved source reveal that the dust emission is not co-spatial with the fullerene emission. Three broad features can be distinguished: a 6--9~\um plateau, a 9--13~\um plateau and the 30~\um feature. The 6--9~\um plateau does not change shape spectrally across the entire field of view, indicating it is produced by stochastically heated material. Spectral variations of the 9--13~\um plateau reveal at least three distinct components, peaking at 9.5~\um, 11~\um and 12~\um respectively. SiC produces most likely at least the 11~\um component and we find evidence that the 12~\um component might be produced by SiC grains too. The variation in relative strength of these components across the field of view hints at dust processing taking place in Tc~1. The 9.5~\um component of the 9--13~\um plateau has a strong correlation with the SiC component(s) and is thus likely produced by similar material. 

\end{abstract}


\section{Introduction}

The planetary nebula (PN, plural PNe) Tc~1 is exceptional because its infrared (IR) spectrum is dominated by ro-vibrational emission bands of fullerenes. Fullerenes are large carbonaceous molecules, constructed from hexagonal and pentagonal arrangements of carbon atoms, shaped like a hollow sphere or ellipsoid. The most stable molecule --- and the archetype --- of this family is \csixty, known as Buckminsterfullerene. Due to their stability against photo-dissociation, fullerenes can survive in the harsh conditions of the interstellar medium (ISM). Indeed, \csixty has been found to be ubiquitous in the ISM, as several diffuse interstellar bands have been identified as \csixtyplus \citep{1993CPL...211..227F, 1994Natur.369..296F, 2015Natur.523..322C}.

Tc~1 is the proto-typical planetary nebula to study C$_{60}$, for two reasons: it does not contain significant emission from PAHs, which usually contaminates some of the IR active bands of C$_{60}$. PAHs produce infrared emission features --- referred to as ``aromatic infrared bands'' (AIBs) --- through vibrational emission; the strongest features are located at 3.3, 5.25, 5.75, 6.2, 7.7, 8.6, 11.2, 12.0, 12.7, and 13.5 \um \citep[e.g.][and references therein]{2008ARA&A..46..289T}. C$_{60}$ produces similar features at 6.5, 7.0, 8.5, 17.4 and 18.9 \um \citep{1990CPL...170..167K, 1991CPL...176..504F, 2016ApJ...826..122B}. We also observe features at 8.8, 12.5, 14.8, 15.6, and 21.8 \um in Tc~1 (Giese et al. submitted), produced by C$_{70}$ \citep{1994CPL...218..295N, 1995CPL...246..321V}. Many of the infrared bands produced by C$_{60}$ and C$_{70}$ overlap with AIBs, thus the lack of AIBs in Tc~1 greatly simplifies the analysis of these fullerene bands. Secondly, Tc~1 is the closest known of such fullerene-dominated nebulae, which makes it easier to study spatial variations. It is resolved for large telescopes, in contrast to a number of other planetary nebulae that show a similar spectrum as Tc~1 but are located in the Magellanic Clouds \citep{2010ApJ...724L..39G,2011ApJ...737L..30G, 2012ApJ...760..107G}. This makes Tc~1 the ideal source to study not only the features of \csixty, but also the physical conditions that give rise to fullerene formation.

Other PNe whose IR spectra are dominated by fullerene emission \citep[M~1-20, M~1-12, K~3-54 and SMP SMC 16;][]{2010ApJ...724L..39G} have IR spectra similar to Tc~1. In addition to the C$_{60}$ IR bands, they show broad plateau-like emission features at 6--9~\um and 9--13~\um and a strong 30~\um feature. The dust continuum emission is also similar in all these PNe. This suggests that a specific environment is required to form fullerenes. The broad emission features are usually attributed to dust species. The 30 \um feature is tentatively attributed to MgS in literature \citep[e.g.][]{1985ApJ...290L..41N, 1985ApJ...290L..35G}, though it is still disputed based on abundance arguments \citep[e.g.][]{2009ApJ...702..680Z}. It is common in many (post)AGB environments \citep{2002ApJ...579..817A} and hence may not be directly related to fullerene chemistry, yet why it is so prevalent in fullerene-rich PNe is unclear.

The two plateaus are incompletely identified at best. No conclusive identification has been offered for the 6--9~\um plateau as of yet. Other fullerene species have been proposed as possible carriers \citep[both smaller and larger than \csixty;][]{2019MNRAS.485.1137C, 2025ApJ...986...77R}; these fullerenes do show emission in the 6--9~\um region. The 9--13~\um plateau on the other hand, has been observed in many (post)AGB environments, and some options have been suggested as carriers. It was first identified as SiC in AGB stars \citep{1974ApJ...188..545T, 1986ApJ...307L..15L, 1997MNRAS.288..431S}, and later also in C-rich PNe \citep{2009ApJ...699.1541B}. Laboratory experiments have shown that SiC indeed produces a feature between 10.5~\um and 12.5~\um \citep{1985A&A...153....1B, 2008A&A...483..661P}, with a shape that is in agreement with the observed 11 \um feature. \cite{2014ApJ...791...28S} identified the 9--13 \um plateau as SiC in a sample of carbon stars, which they suggest might exist as a coating around carbonaceous cores of dust grains. Alternatively, PAH clusters and/or hydrogenated amorphous carbon (HAC) grains have been suggested as an explanation for both the 6--9 \um and 9--13 \um plateau \citep{1990ApJ...365L..23B, 2024A&A...682L..18G}.

We study Tc~1 in detail to investigate in which conditions fullerenes are formed and how these molecules interact with their environment through chemical and physical processes. We present a series of papers ``Studying the fullerene nest'', revolving around JWST observations of Tc~1. J.~Cami et al. (2026, in preparation; hereafter \overviewpaper) provides an overview of our MIRI/MRS observations of Tc~1. M.~Giese et al. (2026, submitted; hereafter \morganpaper) focuses on an in-depth analysis of the fullerene features. In this paper, we study the broad emission features and the continuum emission to characterize the dust component of the environment. 

This paper is structured as follows: we introduce the PN Tc~1 in Section~\ref{sec:tc1} and provide an overview of our observations in Section~\ref{sec:obs}. In Section~\ref{sec:dustfeatures} we investigate the spectral shape and variations of the dust emission. We show the spatial distributions of the dust features in Section~\ref{sec: spatial distributions}. In Section~\ref{sec:discussion} we discuss our findings, including possible carriers for the features, a comparison to optical extinction and dust heating.

\section{Tc~1}
\label{sec:tc1}
\begin{figure*}[t]
    \centering
    \includegraphics[width=\textwidth]{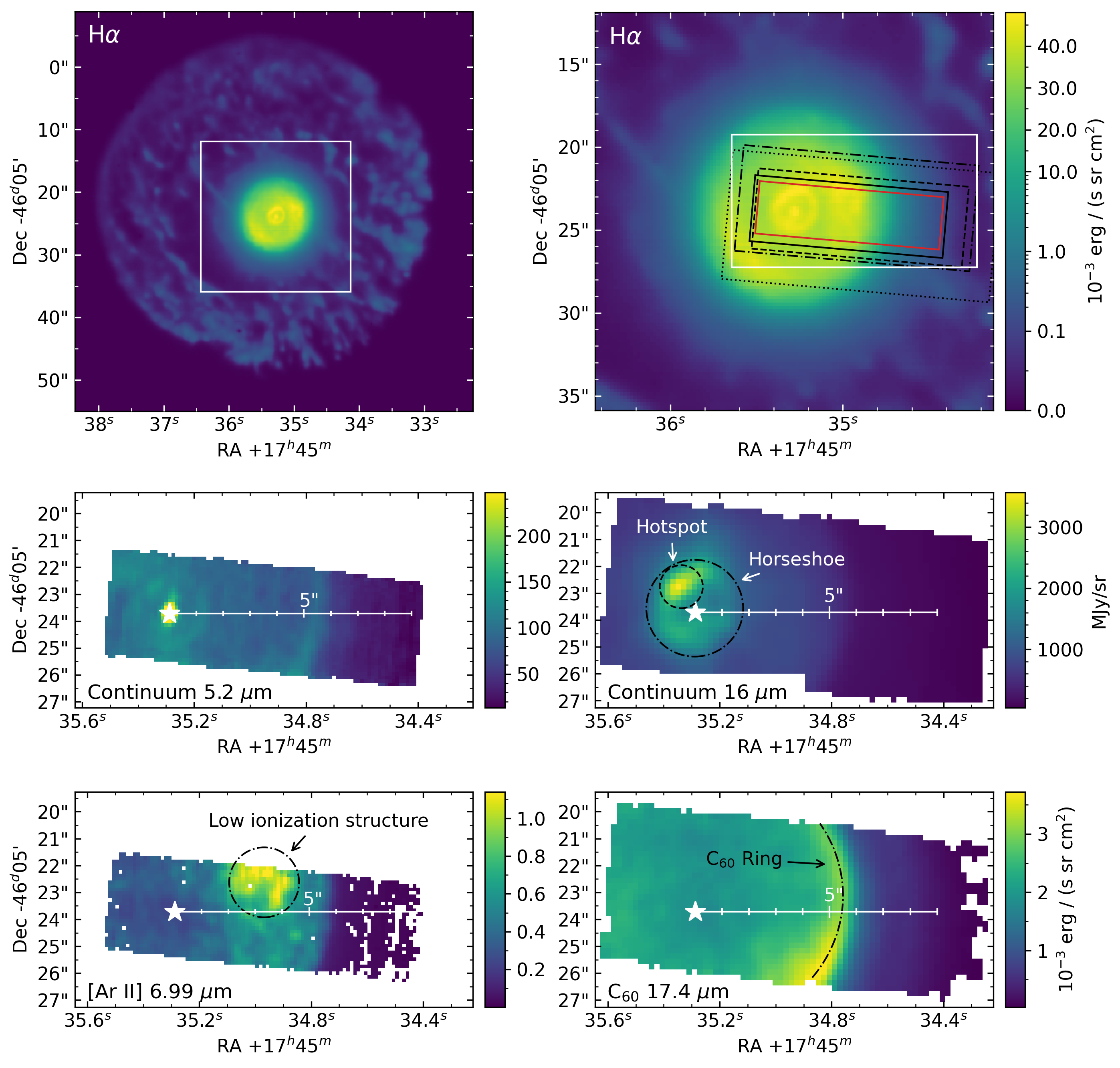}
    \caption{Overview of the structure of Tc 1 (from \overviewpaper). \textit{Top panels:} H$\alpha$ maps from MUSE observations \citep{Walsh:MUSE}. The white box in the left panel shows the FOV of the right panel. The white box in the right panel shows the FOV of the bottom four panels. The FOV of the four MIRI/MRS channels are shown in black. \textit{Middle and bottom panels:} The continuum emission at 5.2 \um and 16 \um, the integrated surface brightness of the [\ArII] line at 6.99 \um and the integrated surface brightness of the C$_{60}$ 17.4 \um band (\morganpaper). Relevant structures are annotated in black. The location of the central star is marked in white, and the horizontal projected distance to the central star is indicated in arcseconds.}
    \label{fig: Tc 1 overview}
\end{figure*}

Tc~1 is a young planetary nebula, with its central star located at $\text{(RA, Dec)} = (17^{\text{h}}45^{\text{m}}35.29^{\text{s}}, -46^{\text{d}}05'23.72'')$ at a distance of
3.53~kpc \citep{Walsh:MUSE}.
In broadband images, Tc~1 appears roughly spherical, consisting of a dense volume with a radius of $\sim$5--6$''$ around the central star (the ``core''), enveloped by a much lower density halo with a diameter of $\sim$55$''$ \citep{Walsh:MUSE}\footnote{At a distance of 3.53~kpc, 1$''$ corresponds to 3530 AU, or 0.017 pc.}. Figure \ref{fig: Tc 1 overview} provides an overview of the structure of Tc~1 (from \overviewpaper). The fullerene emission peaks at the edge of the core, appearing as a thin ring around the central star with a radius of $\sim$5.3$''$. Enhanced continuum emission from dust is found at $\sim$1.2$''$ from the central star. At shorter wavelengths ($\lesssim10$~\um) this emission appears as a ring around the central star; at longer wavelengths it takes the shape of a horseshoe, showing less emission in the west, with a hotspot of emission in the northeast. We refer to this structure as the horseshoe. It is seen not only in the dust emission, but also in many atomic lines (for example in hydrogen recombination lines). Uncorrelated with any other structure, we see enhanced emission of PAHs \citepPAHPaper and low-ionization lines like [\ArII], in a small area around $\text{(RA, Dec)} \approx (+17^{\text{h}} 45^{\text{m}} 35.0^{\text{s}}, -46^{\text{d}}05'22.5'')$; we refer to this as the low-ionization structure (LIS).

\section{Observations and data reduction}
\label{sec:obs}
Tc~1 was observed using the MIRI/MRS \citep{2015Wells, 2023Argyriou} onboard JWST \citep{2023PASP..135f8001G} on August 8, 2024 as part of JWST Cycle 3 program GO-4706 (P.I. Jan Cami). The data cover the wavelength range of 4.9--27.9 \um, utilizing all four channels of the MIRI/MRS instrument (1, 2, 3, 4) and their subchannels (SHORT, MEDIUM, LONG). The data were reduced using the standard JWST pipeline version 1.17.1 \citep{2024Bushouse} with the Calibration Reference Data System jwst\_1322.pmap context file. Further details on the observations and the data reduction are presented in the overview paper (\overviewpaper).

Depending on the feature, different wavelength ranges of the data are needed. Each of the four channels of MIRI/MRS has a different spatial resolution and a different projection of the pixels on the sky, with longer wavelengths having a lower spatial resolution. The three subchannels within each channel have the same spatial resolution, but possibly a slightly different celestial projection, due to small differences in pointing of JWST for subsequent observations. In order to retain the highest possible resolution for the spatial distributions of the dust features, the subchannels are regridded and stitched up to the longest wavelength needed for the feature in question. Stitching is done by scaling the different subchannels with constant factors (i.e. constant in wavelength, but different in each pixel) such that the surface brightness in the overlapping wavelength range between each subchannel is the same. These multiplicative scaling factors are slightly different for each pixel, but generally lie between 0.8 and 1.2. For the 6--9 \um plateau, only data in the first channel are used, the three subchannels of channel 1 are therefore stitched after regridding to channel 1C. The 9--13 \um plateau requires data up to $\sim$15~\um. The subchannels 1A to 3B are regridded to match the celestial projection of channel 3C, and all subchannels of the first three channels are then stitched together. For the continuum and the 30~\um feature, the entire wavelength range is required. The subchannels 1A to 4B are regridded to match the celestial projection of channel 4C, and all subchannels are subsequently stitched together. No PSF matching is performed before stitching.

\section{Dust emission and spectral profiles}
\label{sec:dustfeatures}
\begin{figure}[t]
    \centering
    \includegraphics[width=\linewidth, clip, trim=0.7cm 0.3cm 0.3cm 1.3cm]{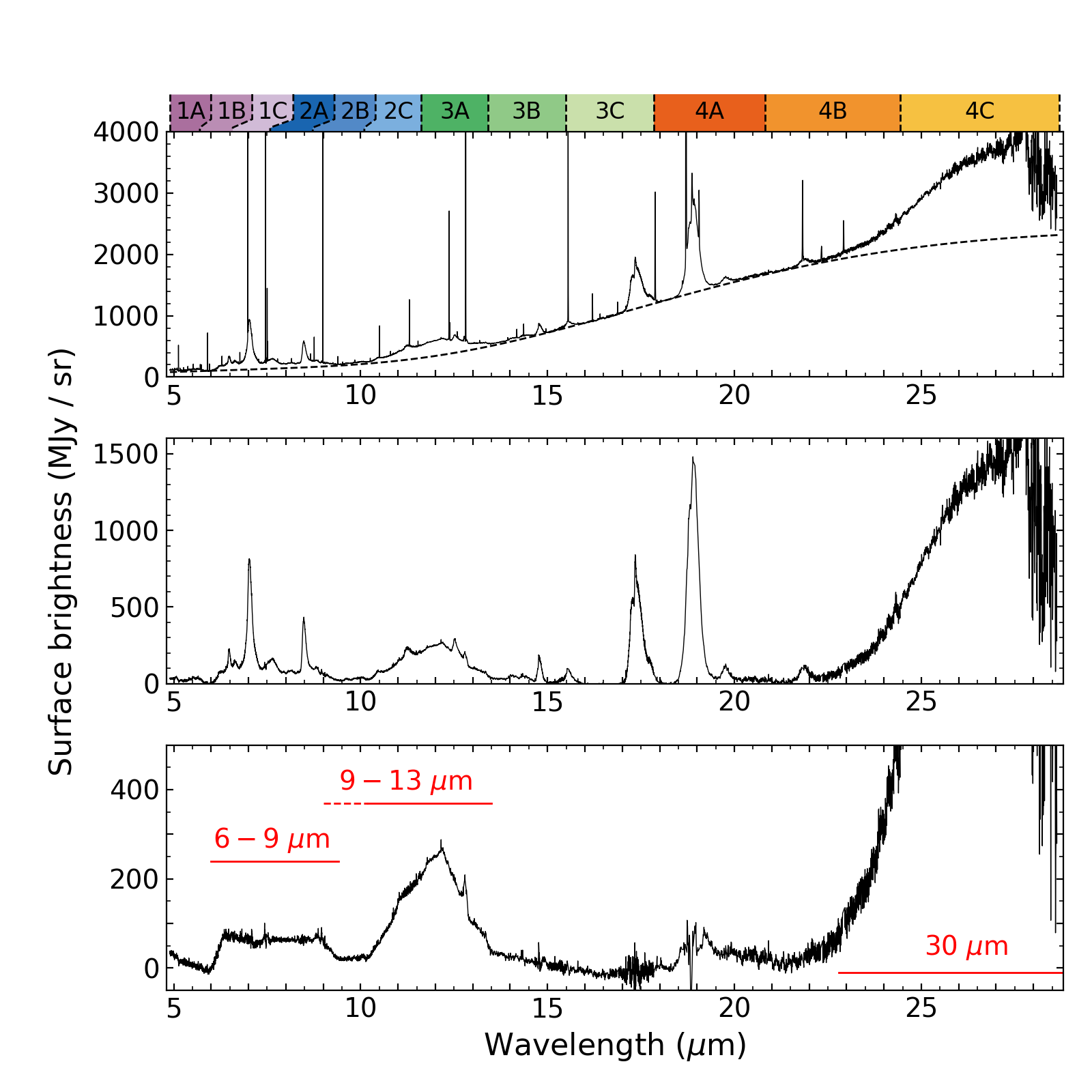}
    \caption{\textit{Top panel:} Typical MIRI/MRS spectrum toward Tc~1 (aperture 5, see Figure \ref{fig:apertures preview}). The dashed line indicates the fitted continuum (see Section \ref{sec: continuum}). The wavelength ranges of the MIRI/MRS subchannels are indicated at the top of the panel. \textit{Middle panel:} The same spectrum as shown in the top panel, but continuum subtracted and with atomic lines removed. \textit{Bottom panel:} The same spectrum as shown in the middle panel, with molecular features subtracted (black). The three plateau-like feature are indicated in red.}
    \label{fig:decomposition example}
\end{figure}

In this section, we investigate the spectral shape and variations of the dust emission in the spectrum of Tc~1. Figure \ref{fig:decomposition example} shows a typical spectrum toward Tc~1. The spectrum consists of four categories of spectral features which can be separated based on their width. In increasing width, one can distinguish (i) unresolved fine-structure and recombination lines, (ii) features identified as C$_{60}$ and C$_{70}$ IR bands, as well as numerous unidentified features --- from here on collectively referred to as ``molecular'' features --- typically 0.3 to 1~\um wide, (iii) three broad features --- two ``plateaus'' and the 30~\um feature, all three generally assumed to be caused by dust species --- multiple \um wide, and (iv) a featureless continuum. First, we present an empirical fitting of the plateaus (Section~\ref{sec: shape of 6-9 and 9-13}), followed by an analysis of the same plateaus using \text{an NMF} based approach (Section~\ref{sec: NMF 9-13 um}).

\begin{figure}[t]
    \centering
    \includegraphics[width=\linewidth, clip, trim=0.1cm 0.2cm 0.8cm 0cm]{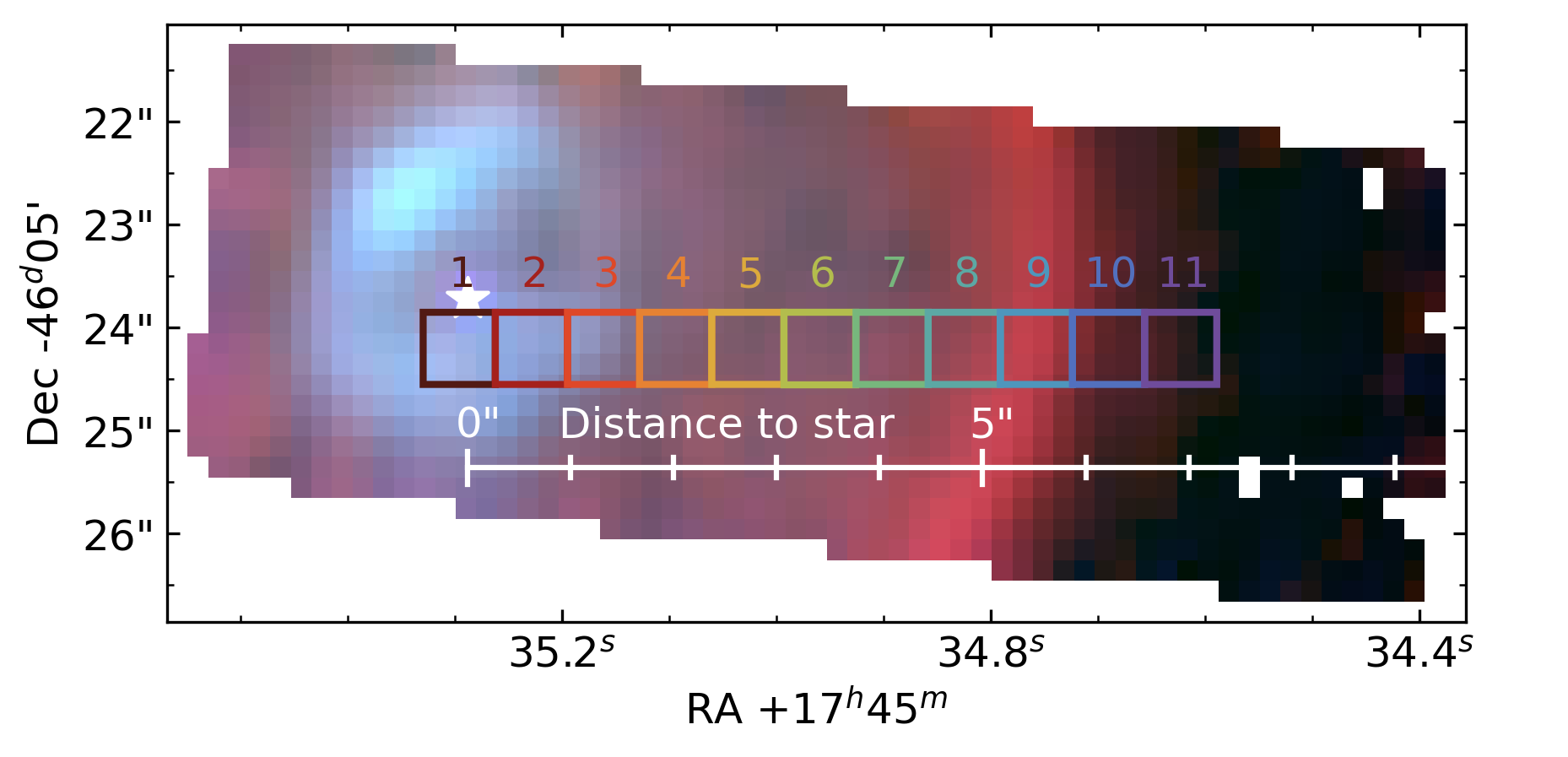}
    \caption{Apertures used for investigating variation in the spectral profiles. The composite image shows the fullerene emission (traced by the integrated surface brightness of the C$_{70}$ 8.8~\um band) in red, the surface brightness of the continuum emission at 13.5--13.9~\um in green and the integrated surface brightness of the 9--13~\um plateau in blue. The central star is indicated in white and the horizontal distance to the star is shown in arcseconds.}
    \label{fig:apertures preview}
\end{figure}

The profiles of the different plateau-like features are examined in different apertures in order to investigate how the shape of the features changes throughout the FOV. Throughout this section, a row of square apertures is considered, as indicated in Figure \ref{fig:apertures preview}. Each aperture is 0.7'' in size; their center coordinates can be found in Table \ref{tab: aperture centers} in Appendix \ref{sec: additional figures and tables}. In order to quantify the variations of the plateaus, the molecular features are fitted and subtracted from the data. A list of the features subtracted from the data is presented in Table \ref{tab: molecular features} in Appendix \ref{sec: appendix molecular features}. This is illustrated in Figure \ref{fig:decomposition example}. Two plateaus can be identified: 6--9 \um and 9--13 \um\footnote{In the example spectrum shown in Figure \ref{fig:decomposition example}, the 9--13 \um plateau only shows emission at 10--13 \um. However, closer to the central star ($\lesssim2''$), significant emission is seen at 9--10 \um. We therefore refer to it as the 9--13 \um plateau, even though the plateau starts at 10 \um in the majority of the FOV.}, and the 30 \um feature, seen from 23 \um onward. The continuum emission is shown with the dashed line; how this continuum is determined is discussed in Section \ref{sec: continuum}.

The surface brightness is typically expressed in units of MJy/sr, which is a frequency based unit. In some cases, it is more convenient to express the surface brightness in the wavelength dependent units \mbox{erg~/~(s cm$^2$ \um sr)}. Throughout this paper, the surface brightness in MJy/sr will be denoted as $F_\nu$, and the surface brightness in wavelength based units will be denoted as $F_\lambda$. The surface brightness in a specific aperture will be referred to as $F_\lambda[a]$, with $a$ the aperture number (1--11, as shown in Figure \ref{fig:apertures preview}).

\begin{figure*}[t]
    \centering
    \includegraphics[width=\textwidth]{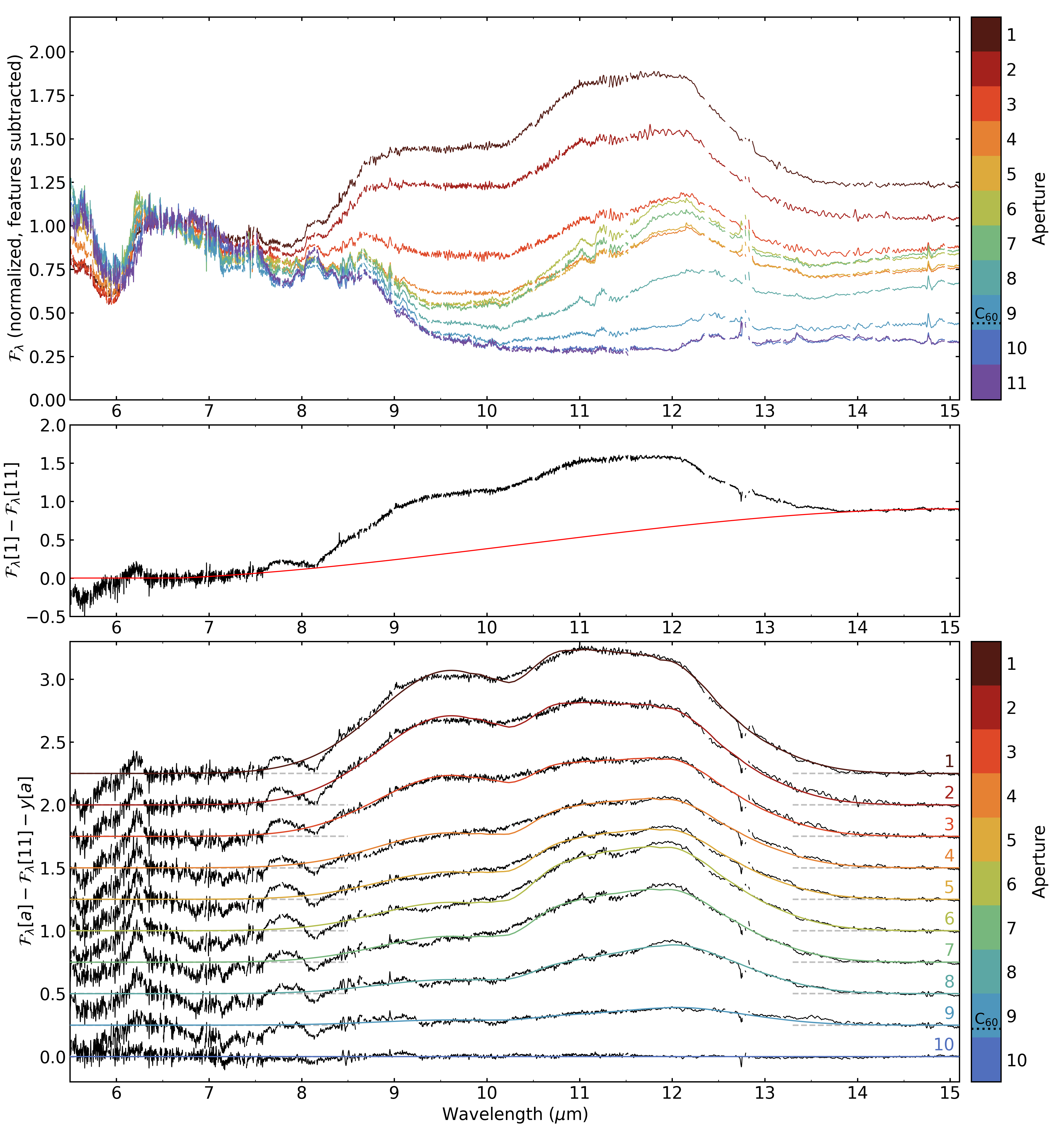}
    \caption{Surface brightness and profiles of the 6--9 and 9--13~\um plateaus in the apertures shown in Figure \ref{fig:apertures preview}. \textit{Top panel:} Surface brightness $\mathcal{F_\lambda}$, in erg / (s cm$^2$ um sr), normalized at 6.5 \um, colored by aperture. Atomic lines have been masked out for clarity. The location of the ring where the fullerene emission peaks is indicated in the color bar by the dotted line. \textit{Middle panel:} The normalized surface brightness of aperture 1 minus that of aperture 11 ($\mathcal{F}_\lambda[1] - \mathcal{F}_\lambda[11]$, black). The difference in continuum emission, $y_\lambda$, is fitted with a third order polynomial, shown in red. \textit{Bottom panel:} The differential surface brightness as shown in the middle panel, now continuum subtracted, for every aperture (black). The different spectra are offset to improve clarity. The offset is indicated by the horizontal gray dashed lines (near 8 and 14~\um). Each profile is fitted with a 3-component model, shown in color (colored by aperture).}
    \label{fig:6-13 plateau profiles}
\end{figure*}

\subsection{The shape of the 6--9 and 9--13 $\mu$m plateaus}\label{sec: shape of 6-9 and 9-13}

The wavelength range from 6 to 13 \um shows a diverse range of shapes throughout the mosaic.  Figure \ref{fig:6-13 plateau profiles} (top panel) shows the surface brightness in each of the apertures. The normalization at 6.5 \um reveals that the 6--9 \um plateau has the same shape for all apertures considered\footnote{The 8--9 \um range does change significantly, but we interpret this as the blue wing of the 9--13 \um plateau overlapping with the red wing of the 6--9 \um plateau.}, while the 9--13 \um plateau changes shape and surface brightness (relative to the 6--9 \um plateau). We will from here on refer to the surface brightness after subtracting the molecular features and normalizing to 1 at 6.5 \um as $\mathcal{F}_\lambda[a]$.

Assuming the spectral shape of the 6--9~\um plateau is intrinsically constant throughout the mosaic, the 9--13~\um plateau can be isolated by subtracting $\mathcal{F}_\lambda[11]$, which shows little to no 9--13~\um plateau emission, from $\mathcal{F}_\lambda[1]$, which has the strongest 9--13~\um plateau emission (Figure~\ref{fig:6-13 plateau profiles}, middle panel).
This differential spectrum shows a smooth profile, starting at $\sim$8.5~\um and ending at $\sim$14~\um, on top of continuum emission. This continuum emission originates from the fact that the dust continuum emission, $F_\lambda^{\text{cont}}$, is slightly different in the two apertures considered.

To extract the 9--13~\um plateau, this difference in continuum emission ($y_\lambda[a] = F_\lambda^{\text{cont}}[a] - F_\lambda^{\text{cont}}[11]$) is approximated by fitting a third order polynomial to the wavelength ranges of 6.4--7.3~\um and 14.5--15~\um. The bottom panel of Figure \ref{fig:6-13 plateau profiles} shows the same differential profile, now continuum subtracted, for each of the apertures ($\mathcal{F}_\lambda^{\text{9-13}}[a] = \mathcal{F}_\lambda[a] - \mathcal{F}_\lambda[11] - y_\lambda[a]$). It can be seen that the 9--13~\um plateau is highly variable and consists of at least two parts: one covering $\sim$8.5--10.5~\um and one covering $\sim$10.5--13~\um. 

\begin{figure}[t]
    \centering
    \includegraphics[width=\linewidth, clip, trim=0.3cm 0cm 1.3cm 0.9cm]{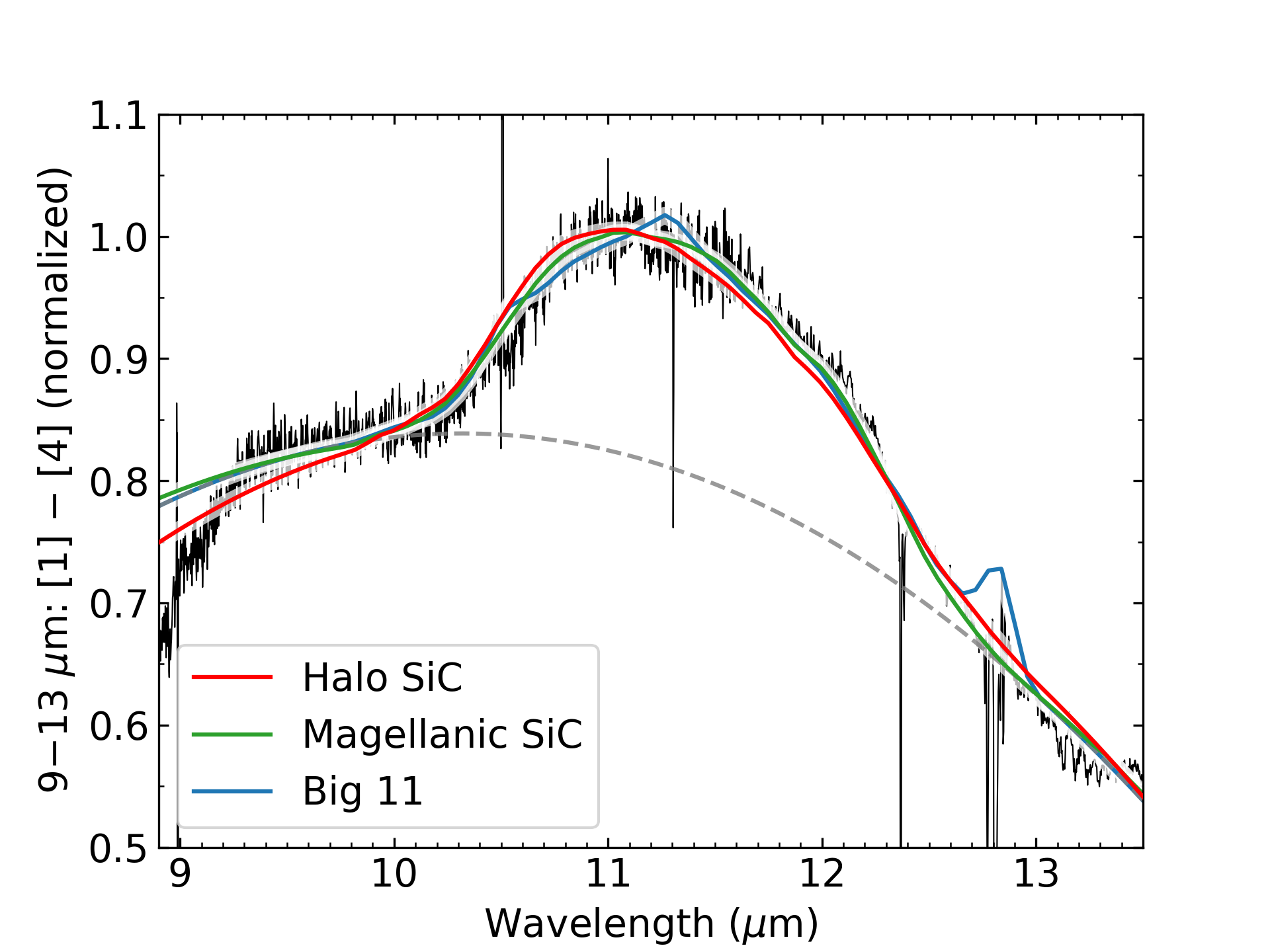}
    \caption{The profile of the 9--13 \um plateau in aperture 1 as shown in Figure \ref{fig:6-13 plateau profiles} (bottom panel), with the profile in aperture 4 subtracted: $\mathcal{F}_{\lambda}$$^{9-13}[1] - \mathcal{F}_{\lambda}$$^{9-13}[4]$ normalized to 1 at 11 \um. Overplotted are three different profiles of SiC, as presented by \cite{2014ApJ...791...28S}, added to a quadratic polynomial (dashed gray line).}
    \label{fig: SiC in 11-13}
\end{figure}

The 10.5--13~\um part itself changes shape throughout the different apertures, and might therefore itself be composed of two or more contributions. For example, the 10.5--13~\um part of the profile of aperture 1 looks convex, while that of aperture 4 looks more triangular. Taking the difference between those two, one can potentially find the contribution that causes this difference in shape (see Figure \ref{fig: SiC in 11-13}). The resulting profile strongly resembles the 11~\um feature seen in AGB stars and C-rich PNe \citep{1974ApJ...188..545T, 1986ApJ...307L..15L, 1997MNRAS.288..431S, 2009ApJ...699.1541B}. 
Laboratory experiments have shown that SiC indeed produces a feature between 10.5~\um and 12.5~\um \citep{1985A&A...153....1B, 2008A&A...483..661P}, with a shape that is in agreement with the observed 11~\um feature. \cite{2014ApJ...791...28S} studied the shape of the 11~\um feature and its variation in a sample of carbon stars. Figure \ref{fig: SiC in 11-13} shows the difference between the 9--13 profile of aperture 1 and 4, and the three SiC spectral profiles presented by \cite{2014ApJ...791...28S}\footnote{We note that the two smaller bumps on top of the ``Big 11'' profile (at 11.3 and 12.8~\um) are molecular features and have been removed from our profile.}. The SiC profiles are added to a quadratic polynomial to take into account the difference of the other contributing features in the 9--13~\um plateau. The SiC profiles are a near-perfect match, indicating that SiC contributes to the 9--13~\um plateau in Tc~1.

\begin{figure*}[t]
    \centering
    \includegraphics[width=\textwidth, clip, trim=0cm 0.3cm 0cm 0cm]{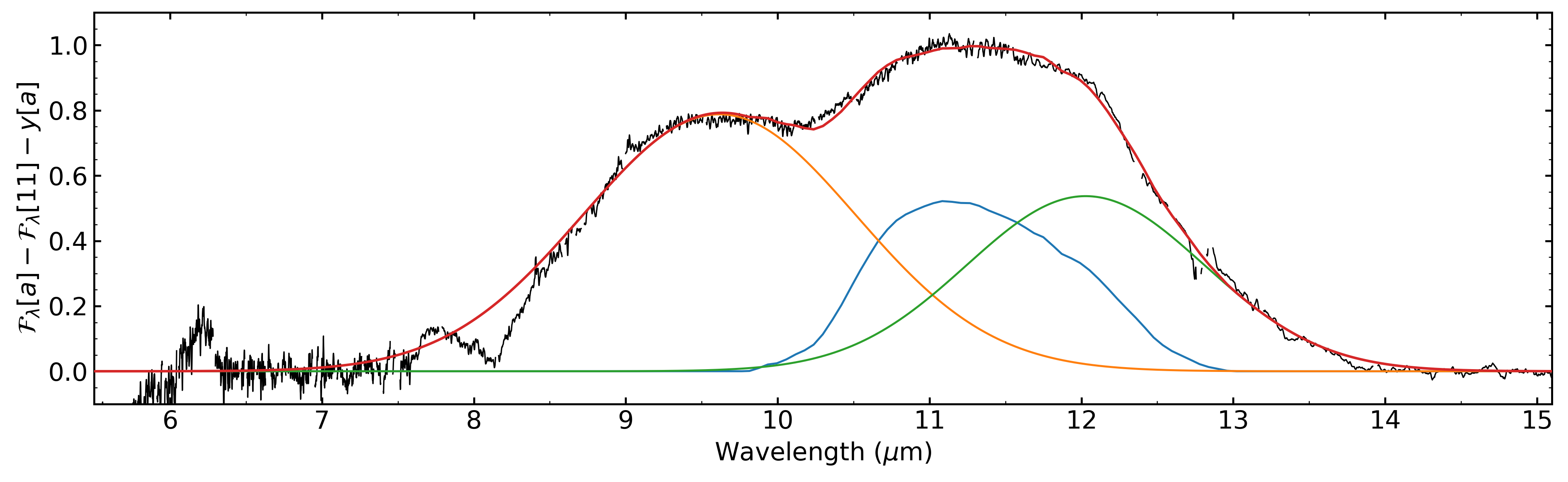}
    \caption{The differential surface brightness as shown in Figure \ref{fig:6-13 plateau profiles} (bottom panel), for aperture 1 (black). The 3-component model is shown in red, with each component separately (blue: SiC, orange: 9.5 \um and green: 12 \um).}
    \label{fig: 6-13 components}
\end{figure*}

Now that we know SiC contributes, we can find what the third component must look like. In addition to SiC, we need to account for the aforementioned feature at $\sim$9.5~\um, leaving a third component at $\sim$12~\um. We therefore fit the 9--13~\um plateau profiles (Figure \ref{fig:6-13 plateau profiles}, bottom panel) with a 3-component model (shown in color) composed of the Halo SiC profile of \cite{2014ApJ...791...28S}, a Gaussian profile peaking at 9.62~\um ($\sigma$ of 0.90~\um) and a second Gaussian profile peaking at 12.03~\um ($\sigma$ of 0.78~\um). Figure \ref{fig: 6-13 components} shows the fit for aperture 1, with each of the components separately. The 9.5~\um component does not appear to be Gaussian in shape, as its blue wing stops abruptly at $\sim$8.2~\um. However, these profiles are the result of multiple steps of subtracting certain features and continua from the data; the exact shape of these profiles should therefore be considered uncertain. Nevertheless, this simple 3-component model fits the extracted 9--13~\um plateau profile fairly well. We estimate the integrated surface brightness of each of the components based on this model (see Section \ref{sec: spatial distributions}). We conclude that the 9--13~\um plateau consists of at least 3 components, peaking at 9.6, 11.2, and 12.03~\um, at least one of which (the 11.2 \um) is due to SiC.

\subsection{The shape of the 9-13 $\mu$m plateau: NMF}\label{sec: NMF 9-13 um}
An alternative, more quantitative method to investigate the different contributions to the 9--13 \um plateau is by performing a principal component analysis (PCA). With a spectrum in each pixel, all $N$ spectra are arranged in an $N \times M$ matrix, with $M$ the number of wavelength elements in each spectrum. This matrix is then factorized into two matrices: $H$, a $C \times M$ matrix containing $C$ ``components'', and $W$, an $N \times C$ matrix containing the weight of each component to reproduce each of the spectra. The components are then optimized to reproduce the variations in the spectra as well as possible. The result is a set of $C$ components, which can be interpreted as template spectra. In the context of the infrared dust emission in Tc~1, one can expect the surface brightness to be a sum of multiple emission features, with an absence of any absorption features. For this reason, it is preferable to perform a non-negative matrix factorization (NMF), which constrains the components and weights to be positive. 

We do not perform the NMF on the surface brightness directly. We isolate the surface brightness of the 9--13 \um plateau using the same process as outlined in Section \ref{sec: shape of 6-9 and 9-13}, producing profiles such as shown in Figure \ref{fig:6-13 plateau profiles} (bottom panel). For the NMF, all pixels in the field of view within $4''$ of the central star are used\footnote{We only use pixels within 4$''$ of the central star because two of the three components are only present within $\sim$2$''$ of the central star (see Section~\ref{sec: spatial maps}). We therefore limit the number of pixels in the training data where only one component is present.}, and we consider only the wavelength range from 8 to 15 \um. Any negative values in the profiles are set to zero, as the NMF requires the input to be strictly positive. Finally, the atomic lines are masked out and the profiles are downsampled in wavelength by a factor of 10, by taking the average of the spectra in bins of 10 wavelength elements, in order to reduce the size of the matrix. This results in an input matrix of 645 spectra, each consisting of 397 wavelength elements.

\begin{figure}[t]
    \centering
    \includegraphics[width=0.999\linewidth, clip, trim=.4cm 0.5cm 0.2cm 0cm]{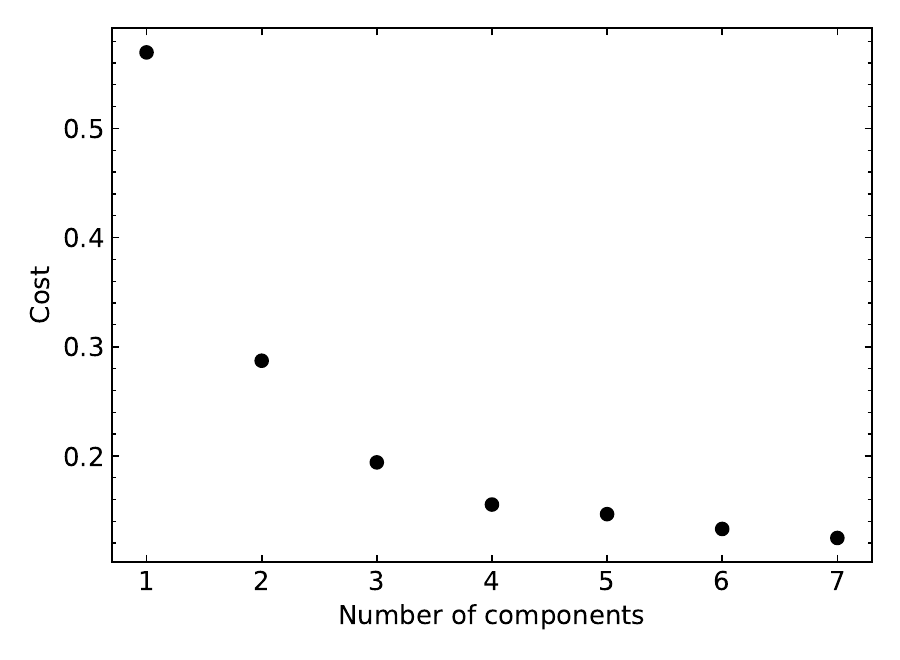}
    \caption{The average of the cost function across the selected field of view for 2 to 9 components.}
    \label{fig: 9-13 nmf n_components}
\end{figure}

We define a cost function to quantify how well a spectrum can be reproduced; the cost for each pixel is equal to the standard deviation of the residuals, and the total cost is the average across all pixels:
$$ c(\boldsymbol{w}_n) = \text{std}\left[ S_{n,m} - (\boldsymbol{w}_n \cdot H)_m \right] $$
$$ \text{Cost}(W) = \frac{1}{N} \sum_{n=1}^N c(\boldsymbol{w}_n) $$
\noindent{}with $\boldsymbol{w}_n$ the weights of the components for pixel $n$ (row $n$ of $W$) and $S_{n,m}$ the spectrum in pixel $n$ at wavelength element $m$\footnote{We use $\mathcal{F}_\lambda^{\text{9-13}}$ in units of erg~/~(s~cm$^2$~\um~sr) here, but the choice of units has little influence on the NMF; using different units mostly scales the cost by a constant.}. Figure \ref{fig: 9-13 nmf n_components} shows the average of this cost function across the field of view --- only including pixels within $4''$ of the central star --- as a function of the number of components. The cost improves significantly from 1 to 3 components, but for more than 3 components the improvement in cost is minimal. Using more components allows to capture more of the variations in the spectra, but it also makes the components more difficult to interpret. Because no a priori knowledge is used to constrain the components, the result of NMF --- and PCA in general --- is a linear combination of the underlying physical contributions to the surface brightness. Using fewer components provides fewer degrees of freedom for such linear combinations, and makes the components generally easier to interpret. We thus conclude that 3 components is the ideal choice. An important limitation of an NMF is that the components are constant, whereas dust emission features can change shape when varying the temperature of the dust grains. This limitation and its implications are discussed in more detail in Section \ref{sec: discussion 9-13 um plateau}.

\begin{figure}[t]
    \centering
    \includegraphics[width=0.999\linewidth, clip, trim=.4cm 0.6cm 0.3cm 0cm]{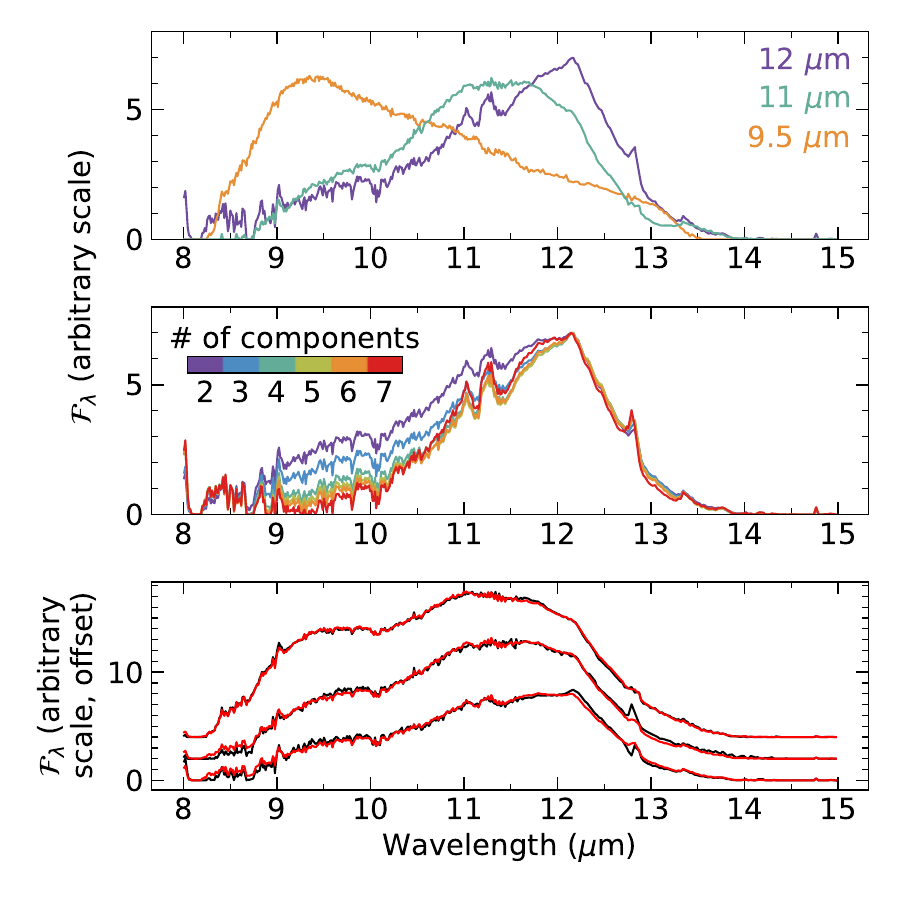}
    \caption{\textit{Top panel:} Components of the NMF, labeled by their approximate peak position. \textit{Middle panel:} The 12~\um component for NMFs using 2 to 7 components. \textit{Bottom panel:} Three examples of reproduced spectra; the surface brightness contribution of the 9--13~\um plateau (black) and the reconstruction using the 3 components (red).}
    \label{fig: 9-13 nmf components}
\end{figure}

The three components of the 9--13 \um plateau produced by the NMF are shown in Figure \ref{fig: 9-13 nmf components} (top panel)\footnote{We note that the amplitudes of the components are arbitrary, as they are degenerate with the scale of the weights: i.e. dividing the amplitude of a component by 2 and doubling the weight of that component everywhere leaves the result unchanged. The relative amplitudes of the components in Figure \ref{fig: 9-13 nmf components} are chosen for readability.}. The teal component (11~\um) strongly resembles the SiC profiles shown in Figure \ref{fig: SiC in 11-13}; this confirms that SiC contributes to the 9--13 \um plateau. 
The purple component peaks at $\sim$12.2 \um (with a shoulder at $\sim 11$ \um) and the orange component peaks at $\sim 9.5$ \um; they correspond to respectively the 12 \um and 9.5 \um components of the simple 3-component model discussed in Section \ref{sec: shape of 6-9 and 9-13}. 

The 12~\um component appears in NMFs for many different choices of the number of components. The other components vary wildly when using more than 4 components, but interestingly, NMFs using 2 to 7 components all produce similar 12~\um components (Figure \ref{fig: 9-13 nmf components}, middle panel). This indicates the 12~\um component is likely closely related to one of the underlying intrinsic contributions to the 9--13~\um plateau. The other two components change more significantly when introducing a fourth component, which might indicate that these are linear combinations of the physical contributions. These NMF components are discussed and interpreted further in section \ref{sec: discussion 9-13 um plateau}, where we discuss possible carriers for the 9--13~\um plateau.

Using these three components, the profiles of the 9--13~\um plateau can be reproduced to great accuracy (with a root-mean-square residual of $\sim5\%$ of the average flux). Figure \ref{fig: 9-13 nmf components} (bottom panel) shows the 9--13~\um plateau in three arbitrary pixels; the reconstructed profiles from the NMF are nearly identical to the original ones. This supports the conclusion drawn from the cost function analysis (Figure \ref{fig: 9-13 nmf n_components}) that three components is sufficient to reproduce the variations in the 9--13~\um plateau.

\subsection{The continuum}\label{sec: continuum}

From an initial inspection of NIRSpec observations of Tc~1 (averaged over the entire NIRSpec FOV), we determine empirically that the continuum emission at wavelengths below 5~\um follows a power law $F_\lambda^{\text{cont}} \sim \lambda^{-0.85}$, or equivalently $F_\nu^{\text{cont}} \sim \lambda^{1.15}$, which we attribute to free-free continuum emission. 
While a more detailed inspection of the continuum emission at wavelengths below 5~\um is deferred to a future paper, this free-free power-law contributes to the continuum emission observed with MIRI/MRS. Therefore, we fit the continuum emission using a two-component model consisting of a modified blackbody surface brightness $B_\nu \lambda^{-2}$, multiplied by an amplitude $A_{\text{bb}}$, and a power law $F_\nu = A_{\text{ff}} \lambda^{1.15}$:
$$ F_\nu^{\text{cont}} = A_{\text{bb}} B_\nu(T_{\text{dust}}) \lambda^{-2} + A_{\text{ff}} \lambda^{1.15} $$

\begin{figure*}[t]
    \centering
    \includegraphics[width=\textwidth]{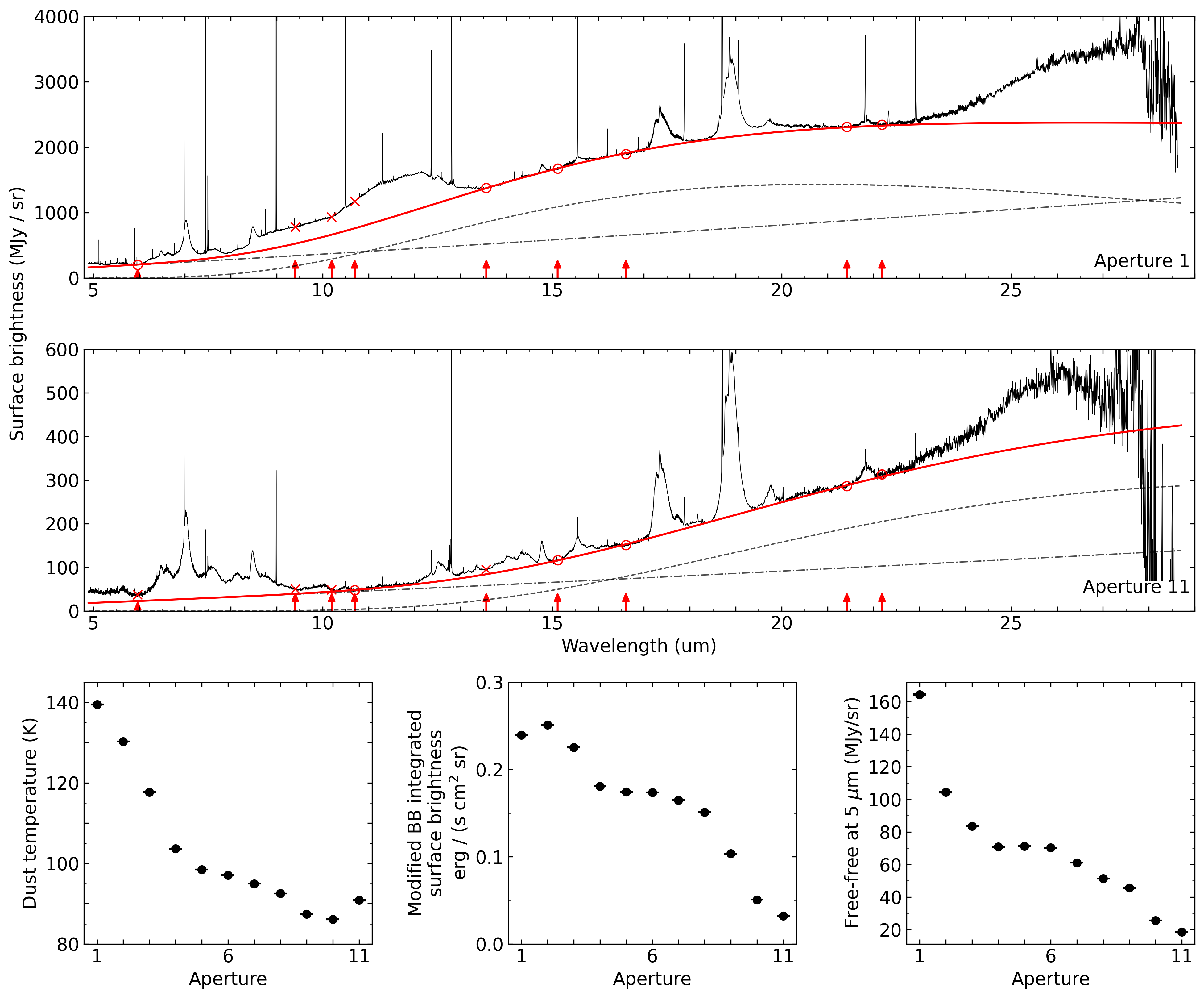}
    \caption{Fits of the continuum emission. \textit{Top two panels:} Surface brightness in aperture 1 (top) and 11 (middle) (black), with the fitted continuum emission (red). The two different components are also shown; the dashed line shows the modified blackbody and the dash-dotted line shows the power law for free-free emission. The anchor points used (hollow circles) and discarded (crosses) for fitting are indicated in red and with the arrows at the bottom of the panel. \textit{Lower panels:} The dust temperature (left), the integrated surface brightness of the modified blackbody (middle) and the amplitude of the free-free emission (right) in each of the apertures. Vertical error bars show the 1-$\sigma$ statistical uncertainties, which are smaller than the dots themselves.}
    \label{fig: continuum fits}
\end{figure*}

Figure \ref{fig: continuum fits} shows the fit of the continuum emission for apertures 1 and 11. The fit considers 9 ``anchor'' points on the spectrum, at wavelengths where --- at least in some areas of the field of view --- no significant emission of any features is present. The first 5 of these anchor points are not always used, depending on the shape of the spectrum. We determine which anchor points are used as follows. The model is fitted with each combination of those 5 first points, for a total of $2^5 - 1 = 31$ possible combinations (the possibility of using none of the first 5 points is not considered, as that would not provide much of a constraint on the free-free component), and the best fit is taken. The best fit is determined by minimizing the $\chi^2$, using the selected anchor points, with an additional term that strongly penalizes discarded anchor points that are below the fitted continuum, effectively providing the constraint that no discarded anchor points can be below the fitted continuum. The same method was used to create the maps of the spatial distributions of the two continuum components (see Section \ref{sec: spatial distributions}).

Beyond the ring where the C$_{60}$ emission peaks (apertures 10 and 11), the surface brightness at 6 \um is significantly above the fitted continuum. At this location, the gas density --- and therefore the electron density --- is much lower, thus one might expect free-free emission to decrease significantly. A different source of continuum emission could be present at the shorter wavelengths, causing the surface brightness at 6 \um to be higher than the fitted two-component model. For this reason, we consider the continuum flux at wavelengths below $\sim$10~\um to be poorly constrained at distances beyond $\sim5''$ from the central star.

Typically, the continuum emission is subtracted from the data to study features like the plateaus. However, the treatment discussed in Section \ref{sec: shape of 6-9 and 9-13} does not require the data to be continuum-subtracted, as the difference in continuum between aperture $a$ and aperture 11 ($y_\lambda[a]$) between 6 \um and 15 \um is approximated by a polynomial. While the continuum fits shown in Figure \ref{fig: continuum fits} provide a good estimate on the global continuum with directly interpretable parameters, they are not sufficient for the purpose of the analysis in Sections \ref{sec: shape of 6-9 and 9-13} and \ref{sec: NMF 9-13 um} without additional corrections to the difference in continuum between aperture $a$ and aperture 11. For this reason the global continuum fit was forgone entirely for the analysis of the 6--9 \um and 9--13 \um plateaus. The fit of the continuum emission is thus only used directly for estimating the 30 \um feature.

\subsection{The 30 $\mu$m feature}\label{sec: 30 um feature}
The blue wing of the 30~\um feature is visible at wavelengths past $\sim 23.5$ \um. The bulk of the 30~\um feature is past the wavelength range of MIRI, which ends at $\sim 28.5$~\um. Nonetheless, the 30~\um feature can be studied from these observations. A few difficulties arise: (i) the continuum emission past the 30~\um feature cannot be constrained from these observations, the analysis of the 30~\um feature thus relies entirely on the estimate of the continuum emission based on the data at shorter wavelengths; (ii) past $\sim$26.5~\um, the signal-to-noise ratio of the data drops dramatically and the shape of the spectrum becomes unreliable \citep{2015PASP..127..686G}.
For these reasons, we expect the analysis of the 30~\um feature to suffer severely from systematic uncertainties. A spectral analysis of the 30~\um feature is not presented in this paper, but the spatial distribution of the integrated surface brightness is discussed in Section \ref{sec: spatial distributions}.

\section{Spatial distributions}\label{sec: spatial distributions}

The spatial distribution of the integrated surface brightness of the various dust features allows us to study the correlation between different features directly. In some cases, however, integrating the plateau is challenging, because the continuum emission is poorly constrained, or, in the case of the 30 \um feature, the feature extends beyond the wavelength range of the data. To circumvent this problem, we opted to use ``proxy'' measures for the integrated surface brightness in these cases, rather than integrating the profiles directly. 

\subsection{Determining the integrated surface brightness}

\begin{figure}[t]
    \centering
    \includegraphics[width=\linewidth, clip, trim=0cm 0cm 0.2cm 0cm]{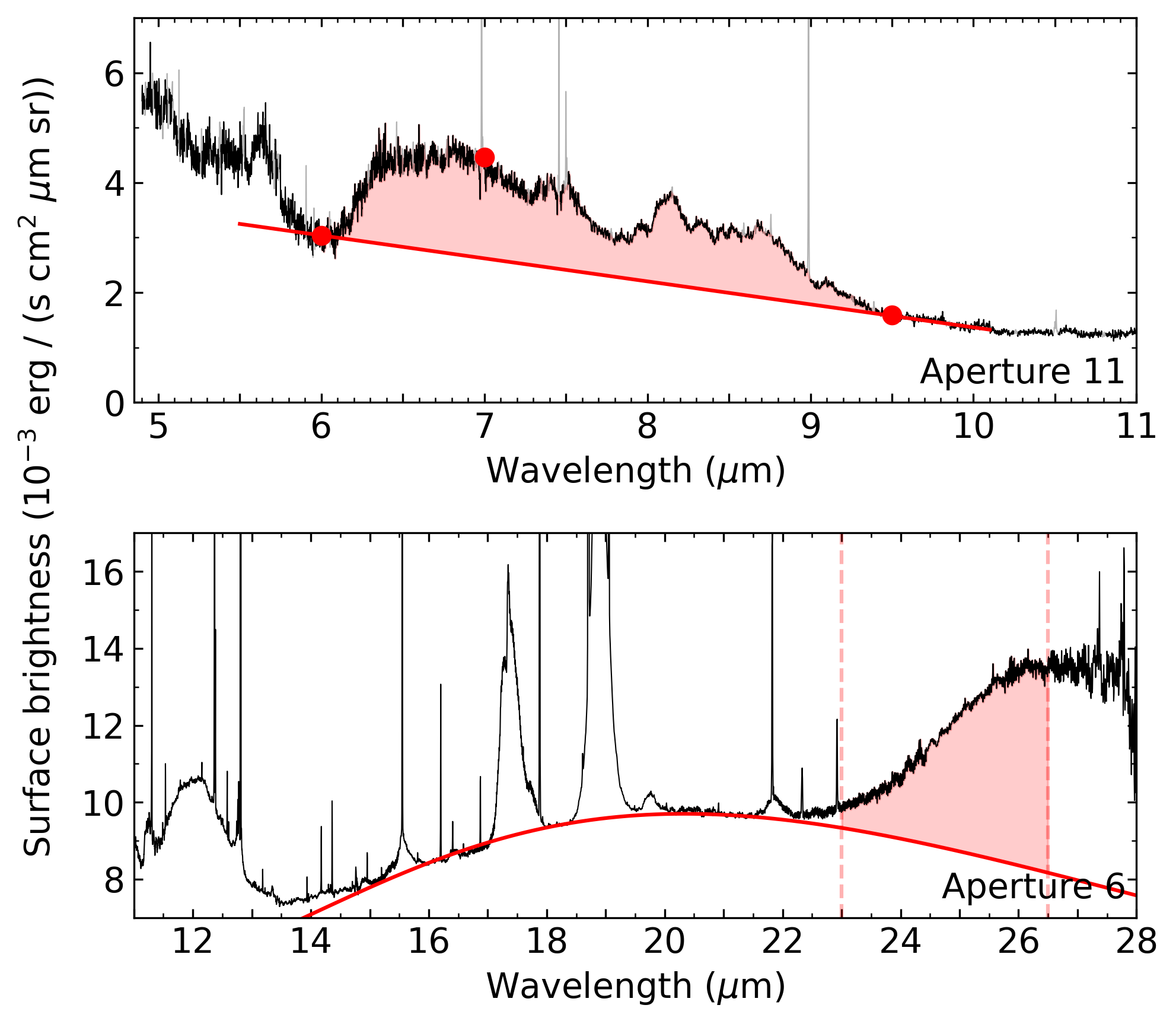}
    \caption{Proxy measures for the integrated surface brightness of the dust features. \textit{Top panel:} The spectrum (black) and the straight continuum between 6 \um and 9.5 \um (solid red).  The 6--9 plateau is measured by the difference in surface brightness between the first two red points, multiplied by 3.62. The factor of 3.62 is determined by integrating the 6--9 \um feature in aperture 11 (red area). Emission lines are masked out for integration (shown in gray). \textit{Bottom panel:} The surface brightness (black) and the continuum fit (solid red). The 30 \um feature is estimated by integrating the flux from 23 \um to 26.5 \um (indicated by dashed red lines), after subtracting the continuum fit (red area).}
    \label{fig: plateau measures}
\end{figure}

For the 6--9 \um plateau, we use the surface brightness increase at the start of the plateau as the proxy. This is illustrated in Figure \ref{fig: plateau measures} (top panel). Specifically, the proxy is the difference in surface brightness between the first two red points, mathematically expressed as:  
$$ F_{6-9 \text{\um}} \approx 3.62 \Bigl[ \langle F_\lambda - F_\lambda^{\text{C}_{60}}\rangle_{7.0\pm0.1 \text{\um}} - \langle F_\lambda \rangle_{6.0\pm0.05 \text{\um}} \Bigr] $$
\noindent{}in units of erg / (s cm$^2$ sr), with $\langle F_\lambda - F_\lambda^{\text{C}_{60}}\rangle_{7.0\pm0.1 \text{\um}}$ the average surface brightness between 6.9 and 7.1 \um, after subtracting the contribution of the C$_{60}$ 7.0 \um band, and $\langle F_\lambda \rangle_{6.0\pm0.05 \text{\um}}$ the average surface brightness between 5.95 and 6.05~\um, both in units of erg~/~(s cm$^2$ \um sr). To convert this proxy measure of the difference in surface brightness between 6 \um and 7 \um to an estimate of the integrated surface brightness, it is multiplied by a conversion factor. The 6--9 \um feature is integrated in aperture 11, where the 9-13~\um plateau is absent, assuming a linear continuum between 6 \um and 9.5 \um, resulting in a factor of 3.62. Using the same conversion factor throughout the FOV implicitly assumes that the shape of the 6--9 \um plateau is the same everywhere, as discussed in Section \ref{sec: shape of 6-9 and 9-13}. The continuum fit discussed in Section \ref{sec: continuum} is not well enough constrained to use it in this integral below $\sim 10$ \um at distances beyond $\sim 5''$, so a linear continuum was used instead.

We follow the method discussed in Section \ref{sec: shape of 6-9 and 9-13} to isolate the 9--13 \um plateau and decompose it using our simple 3 component model. The integrated surface brightness of the 9--13 \um plateau and that of the three separate components is determined from this fit.

Finally, the 30 \um feature is estimated from the 23--26.5 \um wavelength range, as illustrated in Figure \ref{fig: plateau measures}. The fit of the continuum emission (Section \ref{sec: continuum}) is subtracted from the surface brightness, leaving only the contribution of the 30 \um feature beyond 23 \um. Because of the poor data quality past 26.5 \um (see Section \ref{sec: 30 um feature}) we integrate from 23 \um to 26.5 \um. Assuming the shape of the 30 \um feature is constant throughout the field of view, this measure is directly proportional to the total integrated surface brightness of the 30 \um feature.

\subsection{Spatial maps}\label{sec: spatial maps}

\begin{figure}[t]
    \centering
    \includegraphics[width=0.999\linewidth]{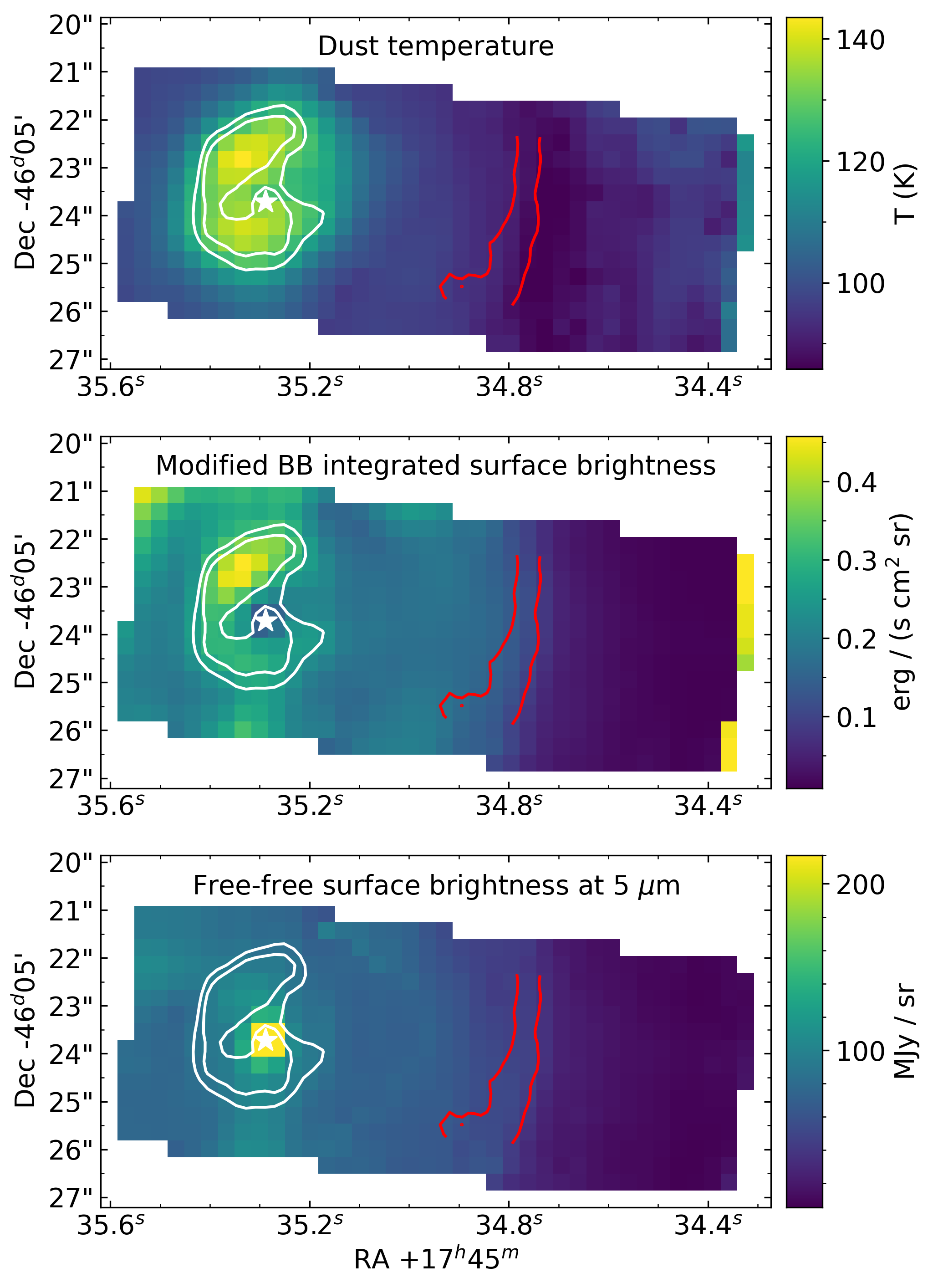}
    \caption{Spatial maps of the dust temperature and the two components of the continuum emission. Contours of the integrated surface brightness of the 9--13 \um plateau are shown in black (Figure \ref{fig: plateau maps}), and contours of the integrated surface brightness of the C$_{60}$ 7.0 \um band are shown in red (\morganpaper).}
    \label{fig: continuum maps}
\end{figure}

Figure \ref{fig: continuum maps} shows the spatial distribution of the dust temperature as estimated from the global continuum fit, the integrated intensity of the modified blackbody component of the continuum, and the amplitude of the free-free component ($A_{\text{ff}}$). The dust temperature is $\sim$120--140~K within $\sim$2$''$ of the central star, with a peak at the hotspot on the northeast side of the horseshoe. The temperature then quickly drops to $\sim$100~K, and stays almost constant up to a distance of $\sim$5$''$. Just past the \csixty ring, the dust temperature reaches a minimum of $\sim$90~K, and rises again up to $\sim$100~K at distances beyond $\sim$6$''$. This is in line with the result for the apertures, shown in Figure \ref{fig: continuum fits} (bottom left).

The integrated surface brightness of the modified blackbody component shows a uniform distribution between $\sim1''$ and $\sim5''$ of the central star. The horseshoe is once again visible, but to a lesser extent than in the spatial distribution of the dust temperature and the 9--13~\um plateau. On the contrary, the hotspot northeast of the central star is much stronger in this map, compared to that of the dust temperature. Beyond $\sim$5$''$, the blackbody emission drops quickly to $\sim$10$\times$ lower values than that observed in the $\sim$1$''$ to $\sim$5$''$ area.

The free-free emission looks just as uniform and beyond the C$_{60}$ ring it drops just like the blackbody component. The central star appears very bright in this map because the continuum of the star is not taken into account. Close to the star, in the peak of the PSF, this causes the free-free emission to be overestimated. As a consequence, the blackbody component is underestimated, which causes the dark spot at the location of the central star in the middle map of Figure \ref{fig: continuum maps}.

\begin{figure}[t]
    \centering
    \includegraphics[width=0.999\linewidth]{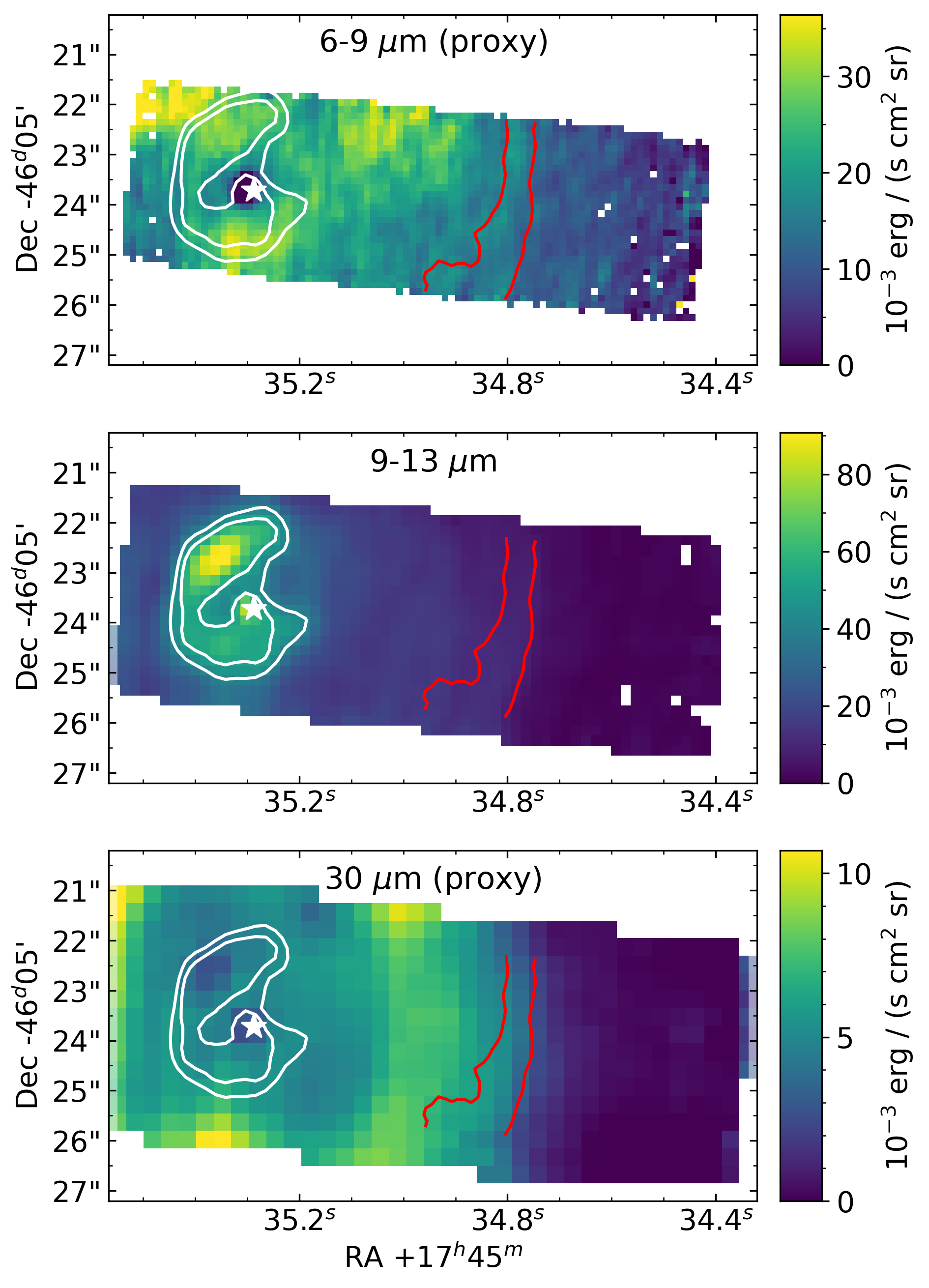}
    \caption{Maps of the integrated surface brightness of the three plateaus. Contours of the integrated surface brightness of the 9--13 \um plateau are shown in white and contours of the integrated surface brightness of the C$_{60}$ 7.0 \um band are shown in red.}
    \label{fig: plateau maps}
\end{figure}

The spatial map of the integrated surface brightness of the 6--9~\um plateau (Figure \ref{fig: plateau maps}) shows fairly smooth emission, generally decreasing with increasing distance to the central star. A small area $\sim$1.5$''$ south of the central star shows slightly increased emission, but the contrast with the emission up to $\sim$5$''$ is no more than a factor of 1.5 to 2. The 6--9~\um plateau also shows a significant amount of emission beyond the \csixty ring.

The spatial distribution of the integrated surface brightness of the 9--13~\um plateau shows rather uniform emission out to the \csixty ring, a strong excess in the horseshoe, and little emission beyond the \csixty ring.

\begin{figure}[t]
    \centering
    \includegraphics[width=0.999\linewidth]{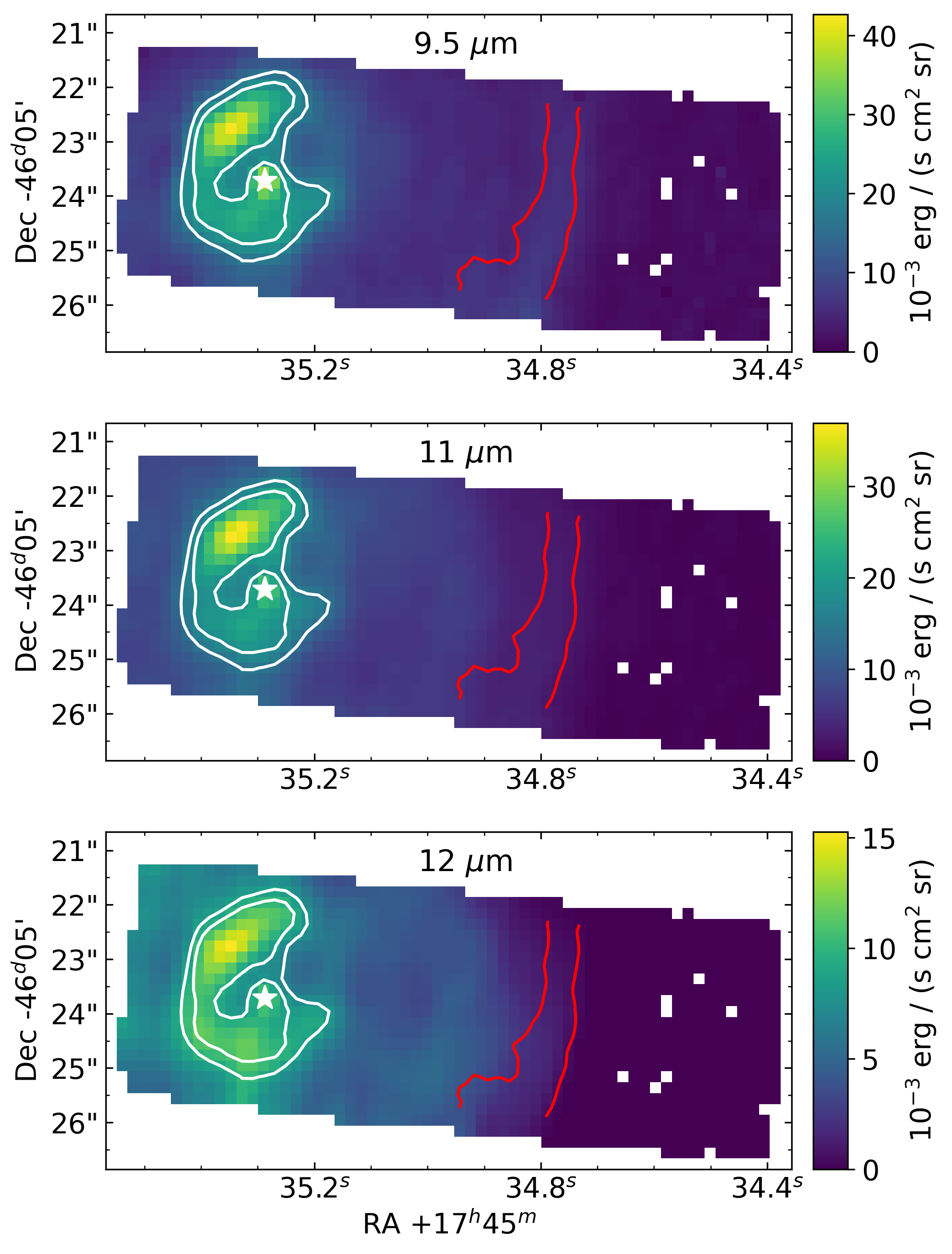}
    \caption{Maps of the integrated surface brightness of each of the three components of the 9--13 \um plateau. Contours of the integrated surface brightness of the 9--13 \um plateau are shown in white and contours of the integrated surface brightness of the C$_{60}$ 7.0 \um band are shown in red.}
    \label{fig: 9-13 components maps}
\end{figure}

Maps of the integrated surface brightness of each of the three components (Section~\ref{sec: shape of 6-9 and 9-13}) are shown in Figure~\ref{fig: 9-13 components maps}. Each component shows the horseshoe around the central star, but the amount of emission at larger distances to the star varies significantly between the three components. The 9.5~\um component shows little emission outside of the horseshoe, at $\sim$1/4 the emission up to the \csixty ring, compared to the horseshoe. The 12~\um component shows emission up to the C$_{60}$ ring at roughly half the intensity of the horseshoe. The 11~\um component sits somewhere in-between, showing about a third of the integrated surface brightness up to the \csixty ring, compared to the horseshoe.

\begin{figure}[t]
    \centering
    \includegraphics[width=\linewidth]{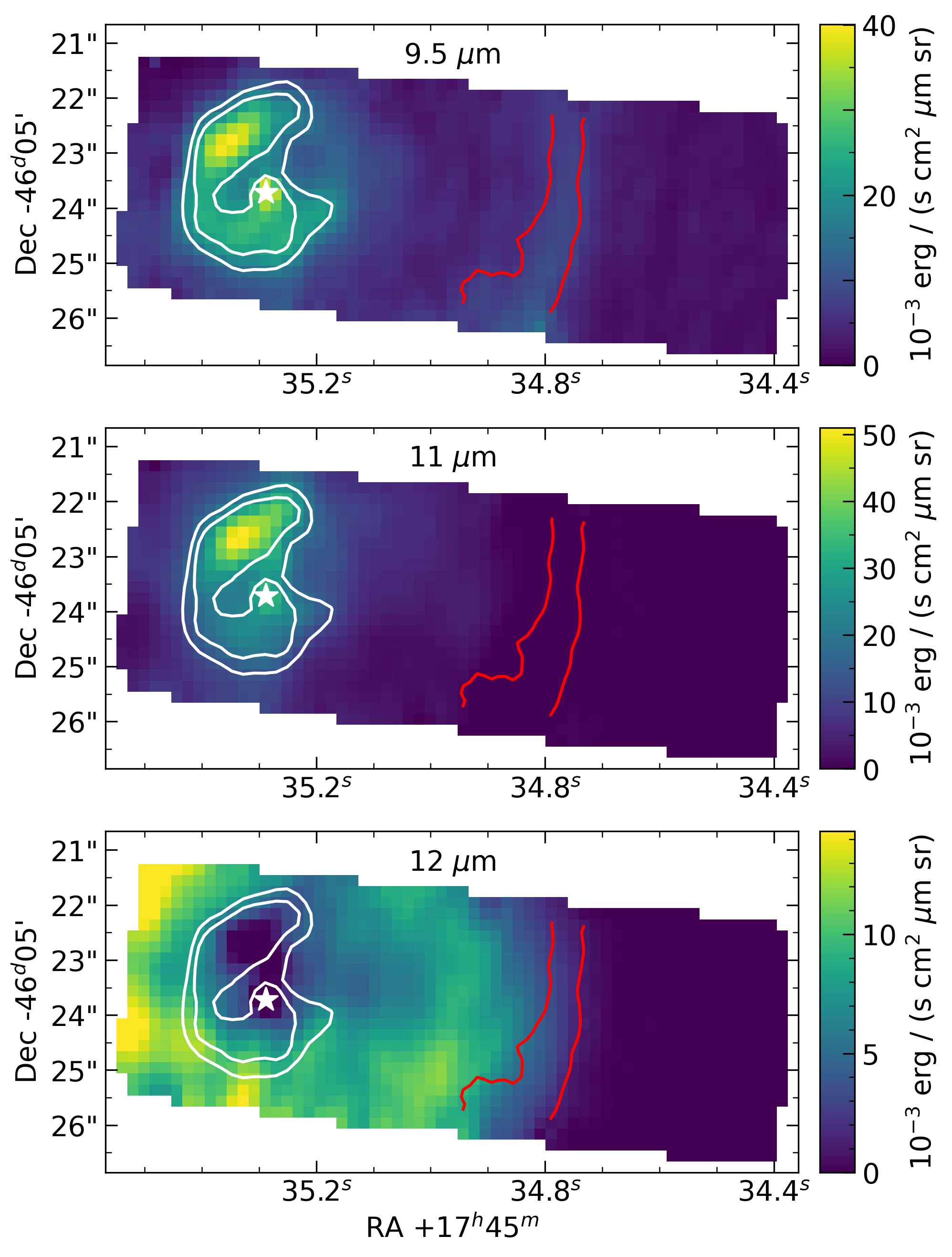}
    \caption{Spatial distributions of the three components of the NMF. Contours of the integrated surface brightness of the 9--13 \um plateau are shown in white and contours of the integrated surface brightness of the C$_{60}$ 7.0 \um band are shown in red.}
    \label{fig: 9-13 nmf spatial distributions}
\end{figure}

The spatial distributions of the three components produced by the NMF of the 9--13~\um plateau are shown in Figure \ref{fig: 9-13 nmf spatial distributions}. The 12~\um component appears quite uniform. This is in agreement with the distribution of the 12~\um component of the simple 3-component model shown in Figure \ref{fig: 9-13 components maps}, which also shows the most emission beyond the horseshoe. Contrary to the spatial distributions simple 3-component model (Figure \ref{fig: 9-13 components maps}), the 12~\um component of the NMF shows no horseshoe --- in fact there is an anti-correlation with the hotspot. The significance of this anti-correlation is unclear, however. The horseshoe is mostly isolated in the 9.5~\um and SiC components of the NMF.

The spatial distribution of the 30 \um feature is uniform up to $\sim$5$''$ (Figure \ref{fig: plateau maps}). A slight anti-correlation with the hotspot can be seen. As discussed earlier in this section, the continuum emission is strongest in the hotspot. Since the 30~\um feature is estimated after subtracting the continuum fit, the anti-correlation with the hotspot might be a result of a systematic uncertainty on the shape of the continuum. A weak ring of enhanced emission is visible at a distance of $\sim$4$''$ from the central star. The same ring can be distinguished in the map of the 12~\um component of the 9--13~\um plateau, albeit to an even lesser degree.

\subsection{Correlations}\label{sec: correlations}

\begin{figure*}[t]
    \centering
    \includegraphics[width=\textwidth, clip, trim=0.5cm 0.5cm 0.5cm 0.5cm]{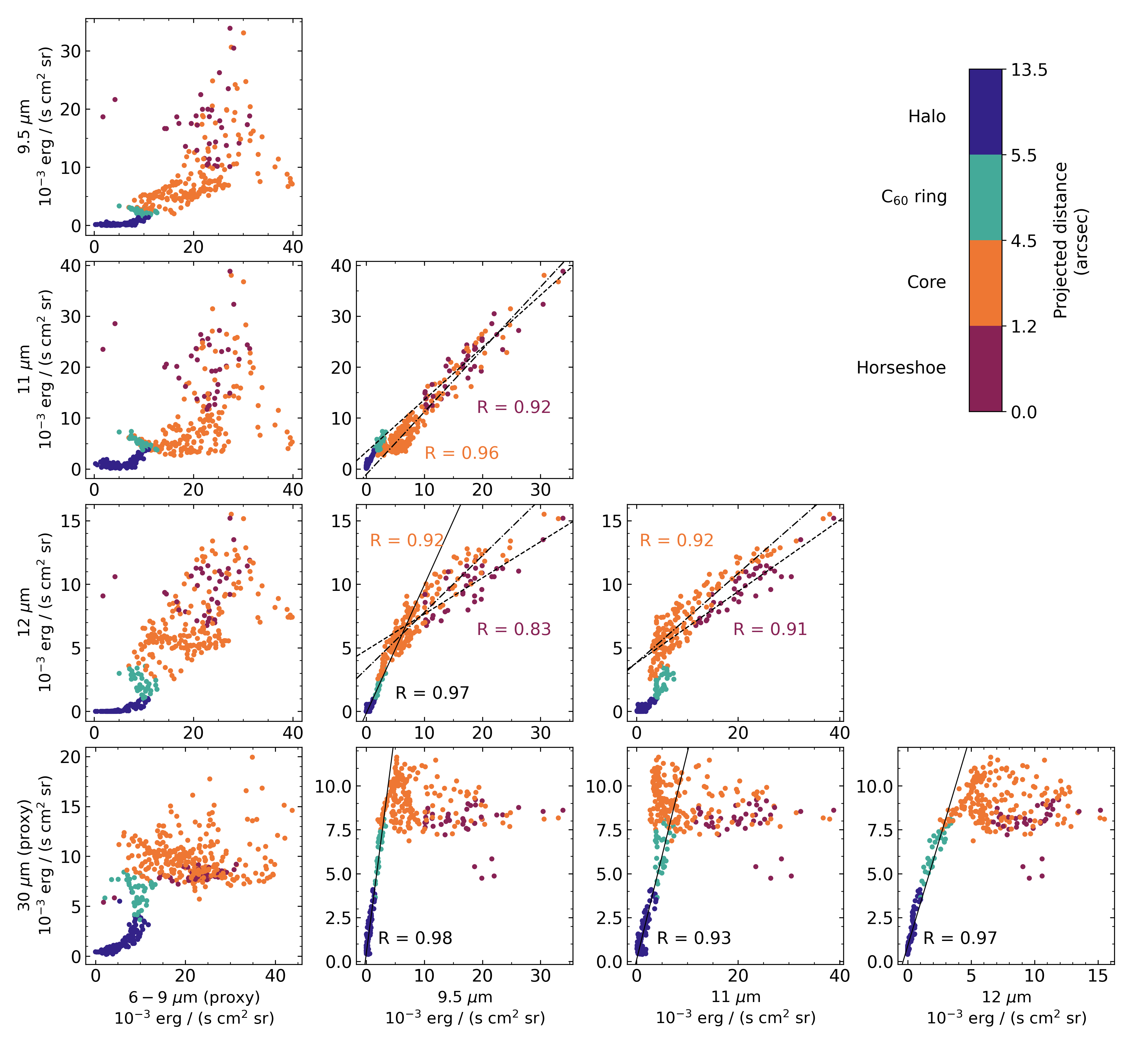}
    \caption{Correlations between the 6--9 \um plateau, the 3 components of the 9--13 \um plateau and the 30 \um feature. The dots are colored by the projected distance of the corresponding pixel to the central star (in arcseconds). A discrete color scale was chosen to represent the major structures of Tc~1: the horseshoe, the \HII region between the horseshoe and the C$_{60}$ ring, the C$_{60}$ ring itself and the halo beyond the C$_{60}$ ring. The most significant correlations are indicated with the black lines. The dashed lines show the correlations in the horseshoe, the dash-dotted lines show the correlations in the \HII region and the solid line shows the correlation in the C$_{60}$ ring and halo. Correlation coefficients using points of only a single region are shown in the same colors as the dots.}
    \label{fig: plateau correlations}
\end{figure*}

With the various plateaus and their components, it can be insightful to look at the correlations between them, shown in Figure \ref{fig: plateau correlations}. For the creation of this figure, all features were fitted on the data, after regridding every channel of MIRI/MRS to the fourth channel. The strongest correlations in the brightest region of the nebula --- the area within $\sim$5$''$ of the central star --- can be seen between the three components of the 9--13~\um plateau. At the C$_{60}$ ring and beyond, all features except for the 6--9~\um plateau show a strong correlation, indicating that the rapid falloff in brightness is likely caused by a drop in density of the dust. This is also supported by the spatial map of the integrated surface brightness of the modified blackbody component of the continuum fit (Figure \ref{fig: continuum maps}, middle panel), which shows the same rapid falloff. The 6--9~\um plateau does not show the same drop beyond the \csixty ring, instead showing a more extended gradient.

\section{Discussion}\label{sec:discussion}

\subsection{The 6--9 $\mu$m plateau}
As mentioned in Section \ref{sec: shape of 6-9 and 9-13}, the shape of the 6--9~\um plateau is constant throughout the field of view, while its strength varies significantly. This suggests the 6--9~\um plateau is likely produced by stochastically heated material. From the dust continuum, we find that the dust temperature varies between $\sim$90--140~K throughout the FOV. Features produced by dust grains would be expected to change shape under such temperature variations. The spatial distribution shows many small-scale variations, which we suspect are at least partially caused by systematic uncertainties or other non-physical effects, likely amplified by our use of a proxy measure for the integrated surface brightness. 
The constant shape of the plateau also suggests it is produced by a single carrier. We consider a few candidate carriers, focusing on carbonaceous species.

We first consider large fullerenes. \cite{2025ApJ...986...77R} present IR emission spectra for various large fullerenes, ranging from 90 to 1750 carbon atoms, computed for absorption of 5--10 eV photons. Their emission spectra show two plateaus, at 6--10 \um and 14--18 \um. The shape of their 6--10 \um plateau matches our observations best for the C$_{90}$ and C$_{162}$ spectra, which show a steep rise at $\sim 6$ \um followed by a gradual decrease from $\sim 6.5$ \um to $\sim 10$ \um. The spectra presented by \cite{2025ApJ...986...77R} show considerable substructure, but the authors stress that they only considered a single conformer per fullerene size, and argue that averaging over many conformers for each size would yield more featureless plateaus. Given the contributions of the many molecular features superimposed on the 6--9 \um plateau, we argue that its exact shape is still unknown, and is not necessarily smooth. 

All fullerene sizes show emission in a second plateau between 14 and 18 \um, which is not observed in our data. While the C$_{90}$ and C$_{162}$ models have the lowest 14--18 \um emission relative to the 6--10~\um emission, this contribution  remains significant. 
\cite{2025ApJ...986...77R} report that for these sizes of fullerenes, higher photon energy leads to a stronger 6--10~\um plateau relative to the 14--18~\um plateau. This is related to the higher excitation energy of the transitions at shorter wavelengths; such higher excitation energy can be obtained by absorption of a higher energy photon or by going to smaller sizes of fullerenes. For a photon energy of 10~eV, the 6--10~\um plateau is roughly twice as strong as the 14--18~\um plateau in their models, up to $\sim 300$ carbon atoms. Unfortunately, Tc~1 --- like all other fullerene-rich PNe --- is a low-excitation object, with a central star effective temperature of $\sim$31,000~K. Given the strength of the 6--9~\um plateau and the radiation field in Tc~1, a significant 14--18~\um plateau would be expected according to the models of \cite{2025ApJ...986...77R}.

The spatial distribution of the integrated surface brightness of the 6--9 \um plateau also may not favor large fullerenes as carriers. The C$_{60}$ and C$_{70}$ emission peaks in a ring around the central star at a distance of $\sim 5''$, while the 6--9 \um emission appears relatively uniform out to $\sim5''$, beyond which it shows an extended gradient toward the edge of the FOV. It is difficult to explain such a markedly different spatial distribution across fullerenes sizes. Nevertheless, their resemblance with the 6--9~\um plateau suggests that species with similar carbonaceous (sub)structures are potential carrier candidates. 

Small fullerenes have also been proposed as the carrier of the 6--9 \um plateau too \citep[e.g.][]{2019MNRAS.485.1137C}. However, in light of our data, this hypothesis seems even less likely than that of the large fullerenes. While small fullerenes do show plateau-like emission at 6--9 \um \citep{2019MNRAS.485.1137C}, they also show emission at 13--17 \um, which we do not observe. The emission at 6--9 \um can be enhanced relative to the 13--17 \um emission at higher photon energies and if the fullerenes are ionized. However, We do not see ionized C$_{60}$ (\morganpaper). Additionally, as with large fullerenes, the very different spatial distribution of the integrated surface brightness between the 6--9 \um plateau and the C$_{60}$ emission does not favor this hypothesis.

A third candidate for the carrier of the 6--9 \um plateau is carbon nanotubes. \cite{2019A&A...631A..54C} present density functional theory (DFT) computed IR emission spectra of nanotubes of various diameters and lengths. Their models show strong emission between $\sim6$ and $\sim10$ \um. The exact shape of the spectrum depends on many variables such as the size of the nanotubes and the structure of their ends. Their spectra show a strong 5.3 \um band associated with the C$\equiv$C symmetric stretching mode of the carbon atoms at the edges of the nanotubes. The same band was found by \cite{2013ApJ...776..102B} who calculated IR spectra of PAHs with some or all hydrogen atoms removed, and detected the 5.3 \um mode in molecules that contain a triple bond between two carbon atoms. We do not detect a  5.3 \um band in Tc~1. However, as it originates from carbon atoms on the edges of the nanotubes, it is absent in closed end nanotubes, which form long capsule-like shapes. \cite{2019A&A...631A..54C} also considered hydrogenated nanotubes, which have hydrogen atoms along both edges. These nanotubes show a strong 3.3 \um band associated with the C$-$H stretching mode and no 5.3 \um band. However, the hydrogenated nanotubes also show much weaker emission in the 6--10 \um wavelength range and a strong 13.0 \um feature that is not seen in Tc~1.

Another candidate for this emission is iron carbide dust. Though, with the carrier of the 6--9 \um plateau likely being stochastically heated material, these dust grains would have to be very small. \cite{liu_preparation_2020} present an IR transmission spectrum of Fe$_5$C$_2$, which shows a broad plateau starting at $\sim 6$ \um and extending to $\sim 12$ \um. A strong feature at $\sim 2.9$ \um is also visible in the transmission spectrum, but the dust temperature would need to be quite high ($\gtrsim 500$ K) for a feature at such short wavelengths to be significant. While the observed plateau in Tc~1 does not extend up to 12~\um (see e.g. aperture 11 in Fig.~\ref{fig:6-13 plateau profiles}), it would be worthwhile to look for a feature at $\sim 2.9$ \um in NIRSpec observations of Tc~1, and investigate the spatial distribution if a feature is found.

Finally, we consider large PAHs as candidate carriers of the 6--9~\um plateau. \citePAHPaper present the aromatic infrared band (AIB) emission in Tc~1, which shows enhanced emission in a low-ionization structure (LIS) northwest of the central star, at (RA,~Dec)~=~(17$^{\text{h}}$45$^{\text{m}}$35.0$^{\text{s}}$,~$-46^{\text{d}}05'22.5''$). Enhanced emission in the 6--9~\um plateau can be seen in the same location (Figure~\ref{fig: plateau maps}). In particular, the 6.2~\um band also shows enhanced emission in the northeast corner of the FOV and in the horseshoe southwest of the star; the 6--9~\um plateau shows these structures too. It is unlikely that our estimate of the 6--9~\um plateau is directly contaminated by the 6.2~\um AIB, as the proxy we use only considers the surface brightness at 6.0~\um and 7.0~\um (Figure \ref{fig: plateau measures}). In fact, if the 6.2~\um AIB had any influence on the proxy for the 6--9~\um plateau, it would cause an underestimate, by contributing to the 6.0~\um surface brightness and thereby decreasing the difference with 7.0~\um. The 6--9~\um plateau thus truly correlates with the 6.2~\um AIB emission.

For large PAHs, the bands between 6--9~\um are indeed expected to blend into a plateau \citep[e.g.][]{2001ApJ...556..501B, Ricca2012, Sidhu2022}.
It may be challenging, however, to explain all the 6--9~\um emission with large PAHs as the AIBs in Tc~1 are very weak---much weaker than the 6--9~\um plateau. An important caveat is that current PAH emission models \citep[e.g.][]{2001ApJ...556..501B, Sidhu2022} assume the PAHs reside in a Photodissociation region (PDR), which is not the case for Tc~1 \citepPAHPaper. Nevertheless, our detection of PAH bands in Tc~1 indicates that PAHs are able to survive in this environment \citepPAHPaper. We therefore propose large PAHs as a promising candidate carrier for the 6--9~\um plateau and await more in-depth modeling of PAHs in \HII regions---to be presented in a forthcoming paper---before drawing a firm conclusion.

The spatial distribution of the 6--9 \um plateau indicates that the nature of the carrier is very different from that of other dust features. The integrated intensity of the 6--9 \um plateau shows an extended gradient beyond the C$_{60}$ ring, out to the edge of the FOV. In contrast, the 9--13 \um plateau and the 30 \um feature show a spatial distribution that resembles the dust continuum emission more closely, with a strong falloff beyond the C$_{60}$ ring. All the candidates for carriers discussed above have potential problems and require further investigation before a more concrete conclusion can be reached. However, given the fullerene-rich nature of Tc~1 and the indication that the carrier of the 6--9 plateau differs markedly from those of the other dust features, we argue that carbonaceous molecules --- whether in the form of large PAHs, large fullerenes or species with similar (sub)structures --- are promising carrier candidates.

\subsection{The 9--13 $\mu$m plateau: SiC}\label{sec: discussion 9-13 um plateau}

The 9--13 \um plateau can be reproduced with three components: SiC, following the profiles of \cite{2014ApJ...791...28S}, a feature at $\sim9.5$ \um and one at $\sim$12 \um. The spatial distribution of these three components is different, indicating they could originate from different carriers. The NMF does show a minor improvement up to four components (Figure \ref{fig: 9-13 nmf n_components}). However, it is important to realize that neither the 3-component model nor the NMF take into account how the shape of the physical contributions vary with temperature. Dust emission is typically parametrized by an emissivity function $Q(\nu)$ multiplied by a blackbody. The variation in shape of the whole 9--13 \um plateau cannot be explained by a temperature variation, as the 9.5 \um component disappears almost entirely beyond $\sim 2''$ from the central star, while the emission at 10--13 \um remains (see Figure \ref{fig: 9-13 components maps}). A blackbody spectrum is much too broad to explain this as a constant $Q(\nu)$ modified by different blackbody spectra. However, we do expect that the variations in shape of the 9--13 \um plateau are partially caused by temperature variations. From the continuum emission we estimate that the dust in the horseshoe is much hotter than the dust in the rest of the ionized region ($\sim$140 K, compared to $\sim$90 K; see Figure \ref{fig: continuum maps}). Secondly, the NMF is performed after a subtraction of the 6--9 \um plateau as well as a continuum fit. The underlying continuum is uncertain and the third order polynomial we use to describe the difference in the continuum between apertures likely does not fully capture the true continuum, leaving additional variation in the training data for the NMF. Most likely the NMF shows minor improvements up to four components because there are additional variations in the shape due to these two reasons.

\begin{figure*}[t]
    \centering
    \includegraphics[width=0.65\linewidth]{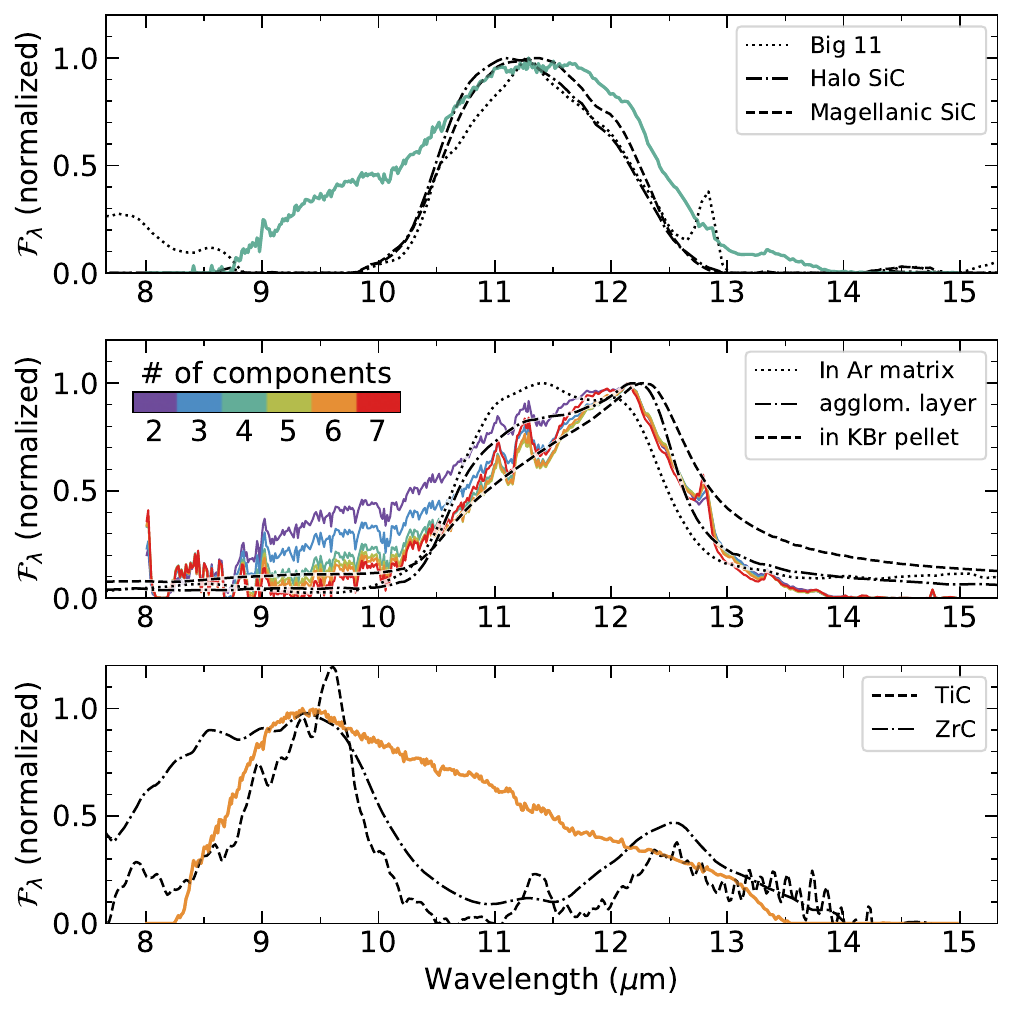}
    \caption{The NMF components discussed in Section \ref{sec: NMF 9-13 um}, compared to promising candidates for the carrier species.}
    \label{fig: nmf identifications}
\end{figure*}

For the remainder of this section, we speculate on the possible carriers of the different components of the 9--13 \um plateau. As discussed in Section \ref{sec: NMF 9-13 um}, one of the components (Figure \ref{fig: 9-13 nmf components}, top panel, teal) resembles the SiC seen in other PNe and post-AGB objects \citep{2008A&A...483..661P, 2009ApJ...699.1541B, 2014ApJ...791...28S}. Figure \ref{fig: nmf identifications} (top panel) shows this component compared to the SiC profiles presented by \cite{2014ApJ...791...28S}. The NMF is performed without any context of physical expectations or any knowledge of the SiC profiles, yet the 10.5--13 \um area of this component matches the SiC profiles almost exactly.

\cite{HENNING2001815} present laboratory spectra of SiC grains with different sizes, shapes and polytypes. They find that the polytype has little influence on the shape of the 10--13 \um feature, but demonstrate that the particle shape, agglomeration state and conductivity introduced by dopants like nitrogen produce significant differences in laboratory spectra. \cite{HENNING2001815} present absorption spectra of a) SiC particles embedded in an argon matrix, b) in the agglomerate layer after evaporation of the matrix and c) those same particles embedded in a KBr pellet. The latter two produce a profile that peaks around 12 \um, with a shoulder around 11 \um. These profiles look similar to the 12 \um component of the NMF discussed in Section \ref{sec: NMF 9-13 um}. Figure \ref{fig: nmf identifications} (middle panel) shows the 12 \um component for NMFs with various numbers of components, with the three profiles of \cite{HENNING2001815} overplotted. The profiles of the SiC grains in the agglomerate layer and KBr pellets closely match the 12 \um component, within the variation of the shape depending on the total number of components. We therefore propose the 12 \um component might also be produced by SiC grains; perhaps of a different size, shape or with the inclusion/exclusion of some dopants on the surface of the grain, compared to the grains producing the $\sim$11 \um component. The different spatial distribution of the two SiC components might then hint at dust processing taking place in the nebula.

Other carbides could be considered as carriers for the 9.5~\um component, though the options are limited due to abundance considerations. Since the 9.5~\um component is quite strong, the carrier must be relatively abundant. Recall that the 9.5~\um component shows an excellent correlation with the SiC component (Section \ref{sec: correlations}, Figure \ref{fig: plateau correlations}), which does suggest that the two are related. One promising candidate is zirconium carbide (ZrC). Given the evolutionary history of Tc~1, one can expect a strong enrichment in s-process elements. Indeed, an abundance analysis of optical emission lines shows an enrichment in Kr by 0.8 dex from the solar abundance, based on the [\KrIII] 6827 {\AA} line \citep{2019MNRAS.490.2475A}. In our MIRI observations, we measure a Kr enhancement of a factor of $\sim$4.6 relative to O ($[\text{Kr/O}] = 0.66 \pm 0.17$) (\overviewpaper). According to \cite{2008ApJS..174..158S}, such Kr enrichment can be produced by s-process nucleosynthesis in the progenitor star. Zirconium could easily be enhanced by an order of magnitude during the third dredge-up in the AGB phase \citep{2016ApJ...825...26K, 2021ApJ...908...55B}. Additionally, ZrC has a higher condensation temperature than SiC. In fact, it has a higher condensation temperature than any other carbide detected in space \citep{1995Metic..30..661L}. However, with the solar abundance of Zr being 5 orders of magnitude down from Si, it may be challenging to create a feature as strong as the 9.5~\um component, even with a strong Zr enrichment. Much more abundant than Zr is Ti, with the solar abundance more than 2 orders of magnitude higher. TiC also has a very high condensation temperature \citep{1995Metic..30..661L}.

Unfortunately, few laboratory infrared spectra of TiC and ZrC are available in literature; \cite{2003A&A...409..159K} present an absorption spectrum of ZrC presolar grain analogs. The spectrum shows a broad feature peaking at 9.4~\um. This resembles the 9.5~\um component in the NMF (see Figure \ref{fig: nmf identifications}, bottom panel). The profile of ZrC shown in Figure \ref{fig: nmf identifications} was adapted from \cite{2003A&A...409..159K}; the continuum was fitted with a polynomial and subtracted. The resemblance is not nearly as strong as for the SiC profiles; in the laboratory spectrum presented by \cite{2003A&A...409..159K}, the 9.4~\um component extends down to below 8~\um, while the 9.5~\um component of the NMF ends abruptly at $\sim$8.5~\um. Their spectrum also shows a feature at 12.5~\um, but little absorption between 10 and 12~\um. \cite{2003MNRAS.343..385K} present absorption spectra of TiC presolar grain analogs. The profile of TiC shown in Figure \ref{fig: nmf identifications} was adapted from \cite{2003MNRAS.343..385K} using a similar approach as for the ZrC profile. The spectrum of TiC also shows a strong component at 9.5~\um; this component is narrower than that of ZrC and follows the blue edge of the NMF component more closely. Just like ZrC, TiC show a weaker feature at 12.5~\um, but little absorption between 10 and 12~\um. We tentatively propose TiC and ZrC as a promising candidates for the 9.5~\um component of the 9--13~\um plateau, but more laboratory measurements are required to draw any conclusions.

An alternative candidate for the carrier of the 9--13~\um plateau in carbon-rich objects like Tc~1 is hydrogenated amorphous carbon (HAC) dust. \cite{2024A&A...682L..18G} used laboratory optical constants of HAC-like dust at $\sim$300 K and the radiative transfer code {\sc Cloudy} to model the 9--13~\um plateau in Tc~1, as seen by Spitzer. They reproduce a feature at 11--12.5~\um, but their model does not reproduce the entire 9--13~\um plateau. Dust models such as those presented by \cite{2024A&A...682L..18G} have numerous free parameters (e.g. the distribution of grain sizes and shapes, and the degree of aromaticity), which can influence the predicted emission spectrum significantly. In this context, the different spatial distribution of the 9--13~\um components would imply different degrees of HAC processing along the nebula. However, given how well two of the three components of the 9--13~\um plateau are described by SiC, we favor carbides as carriers for the 9--13~\um plateau. Nevertheless, given the carbon-rich nature of Tc~1, it is a reasonable suggestion that HAC-like dust makes up a significant fraction of the dust. We suggest that if HAC dust is present in Tc~1, it most likely contributes to the continuum, rather than one of the plateau features.

\subsection{The 15--20 $\mu$m continuum and 19.5 $\mu$m feature}

\begin{figure}[t]
    \centering
    \includegraphics[width=0.999\linewidth]{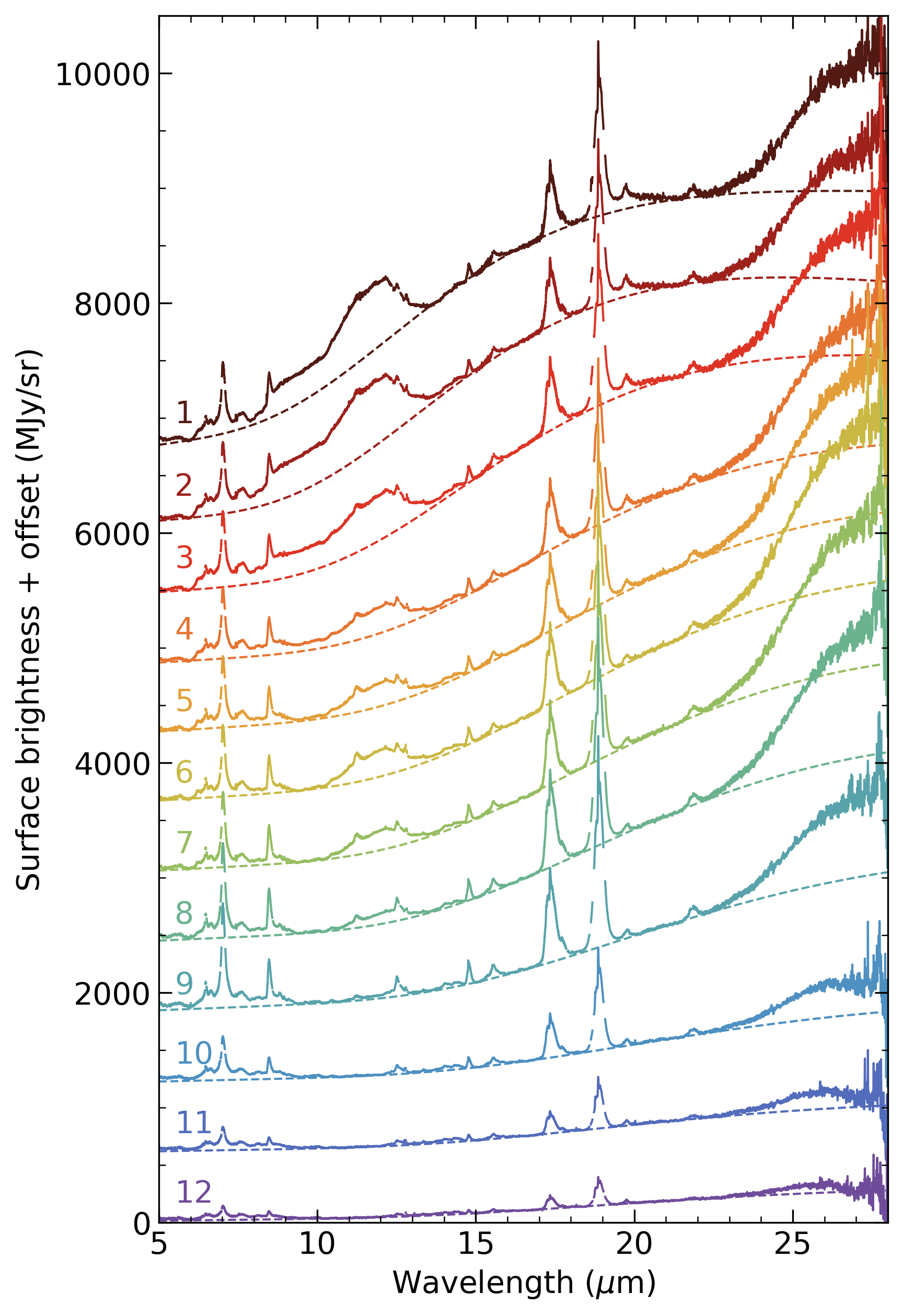}
    \caption{Surface brightness in apertures 1--12 (solid), with continuum (dashed), offset for clarity.}
    \label{fig: 15-20 with continuum}
\end{figure}

\begin{figure}[t]
    \centering
    \includegraphics[width=0.999\linewidth]{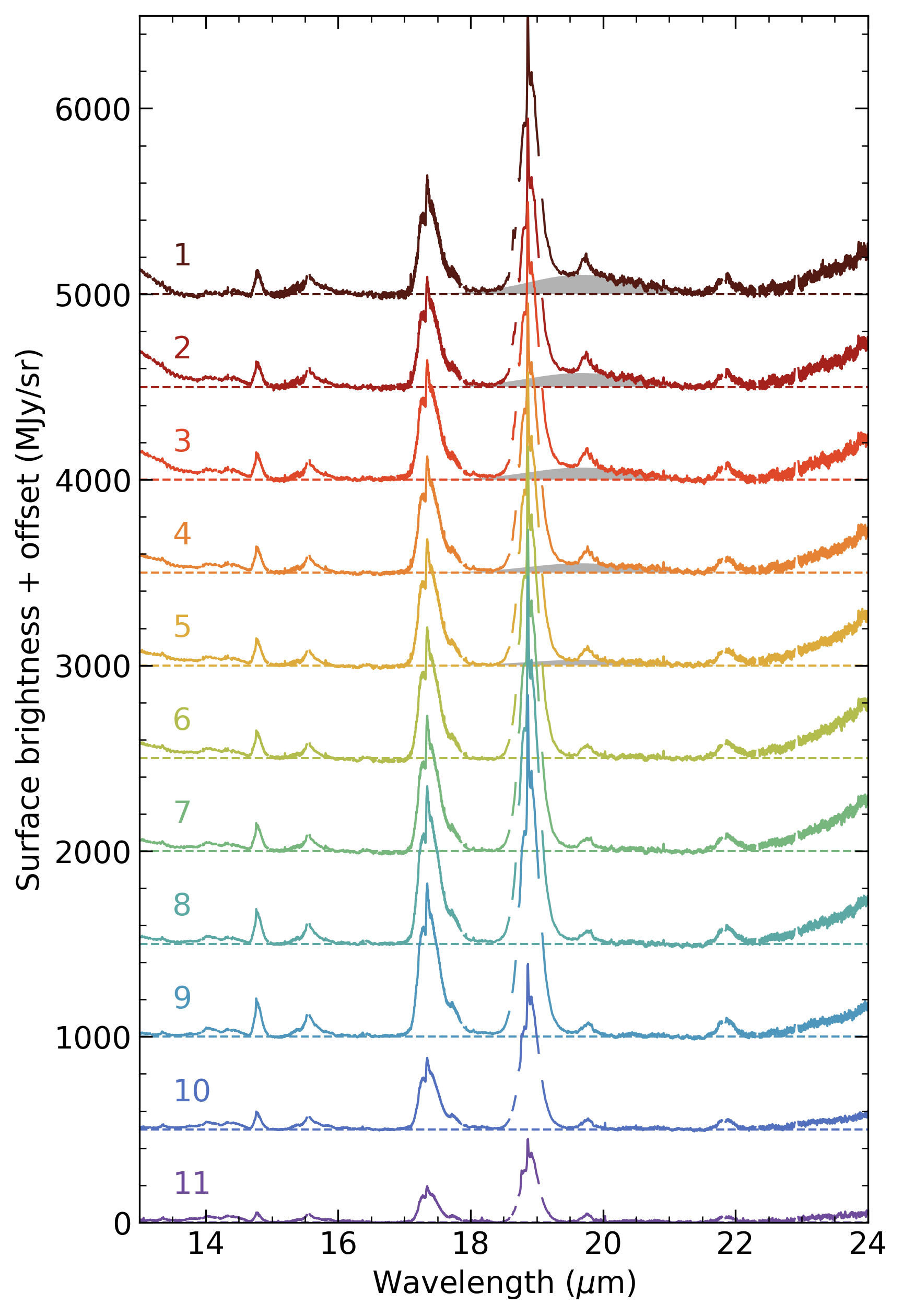}
    \caption{Surface brightness in apertures 1--12 (solid), continuum subtracted, offset for clarity. The 19.5~\um feature is indicated in gray for apertures 1--5.}
    \label{fig: 15-20 continuum subtracted}
\end{figure}

\begin{figure}[t]
    \centering
    \includegraphics[width=0.999\linewidth]{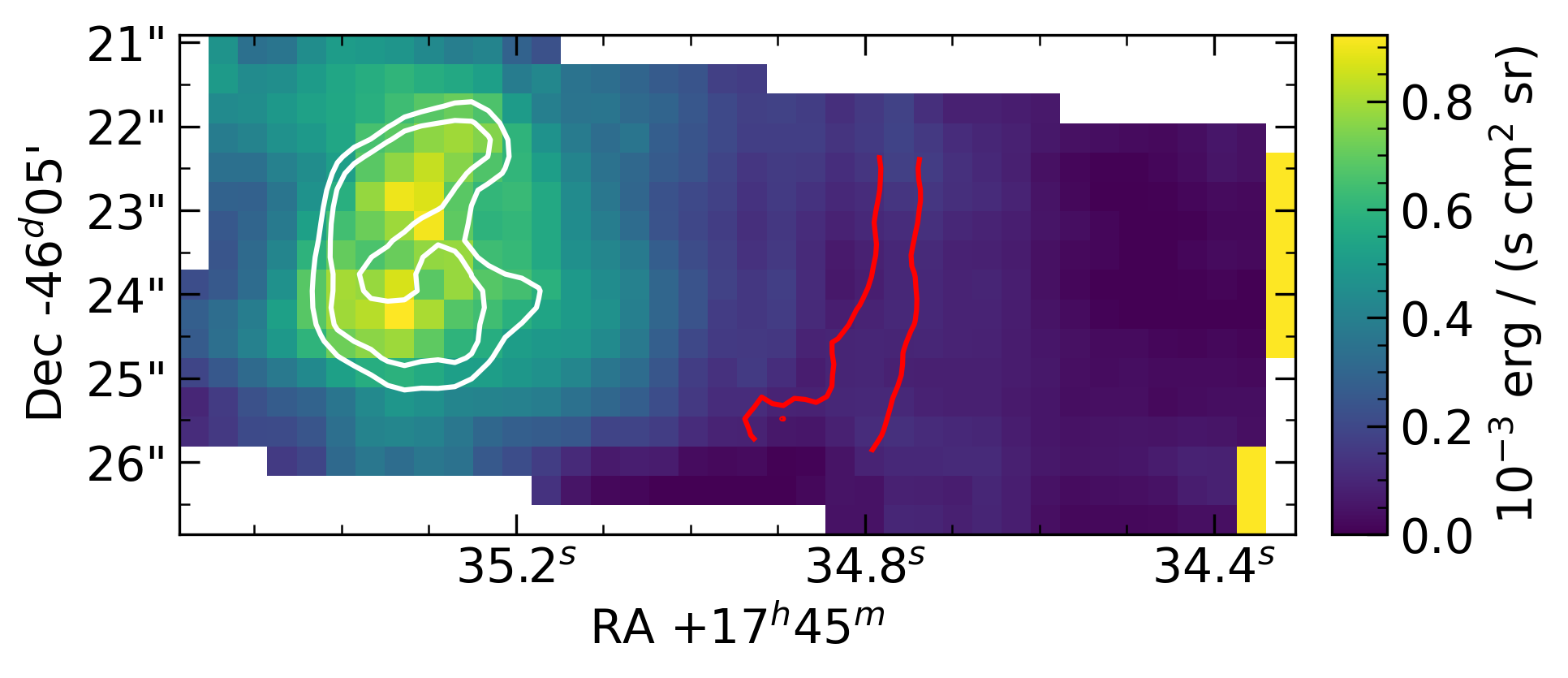}
    \caption{Integrated surface brightness of the 19.5~\um feature. Contours of the integrated surface brightness of the 9--13~\um plateau are shown in white and contours of the integrated surface brightness of the C$_{60}$ 7.0~\um band are shown in red.}
    \label{fig: 19.5 um image}
\end{figure}

The 15--20~\um region shows a clear bump in apertures near the star (Figure \ref{fig: 15-20 with continuum}, apertures 1--3). We interpret this bump as increased emission from warm dust since it can be captured by our continuum fit using a single modified blackbody profile. However, the lack of spatially resolved far-IR data currently makes it impossible to constrain the continuum beyond the 30~\um feature, so we cannot verify whether this choice of continuum is correct. After subtracting this continuum, a weak feature can be discerned between 18--21~\um in apertures 1--5 (Figure \ref{fig: 15-20 continuum subtracted}). This feature has not previously been seen in carbon-rich PNe or post-AGB objects. Fitting a Gaussian profile centered on 19.5~\um to the continuum-subtracted surface brightness in aperture 1, we find a best-fitting FWHM of $\sim$1.9~\um ($\sigma=0.8$~\um). We then fix the standard deviation of the profile to 0.8~\um and fit the amplitude of the 19.5~\um feature in each pixel to derive the spatial distribution shown in Figure \ref{fig: 19.5 um image}. As indicated by the continuum-subtracted surface brightness in the apertures (Figure \ref{fig: 15-20 continuum subtracted}), this feature is strongest in the horseshoe and drops to effectively zero beyond $\sim$3$''$ from the central star. Given this spatial distribution, we propose that the 19.5~\um feature may be related to one of the dust species responsible for the strong emission in the horseshoe, perhaps the carrier of the 9.5~\um component of the 9--13~\um plateau.

\subsection{Comparison to optical extinction}
\begin{figure}[t]
    \centering
    \includegraphics[width=\linewidth, clip, trim=1.3cm 0.3cm 0.4cm 0cm]{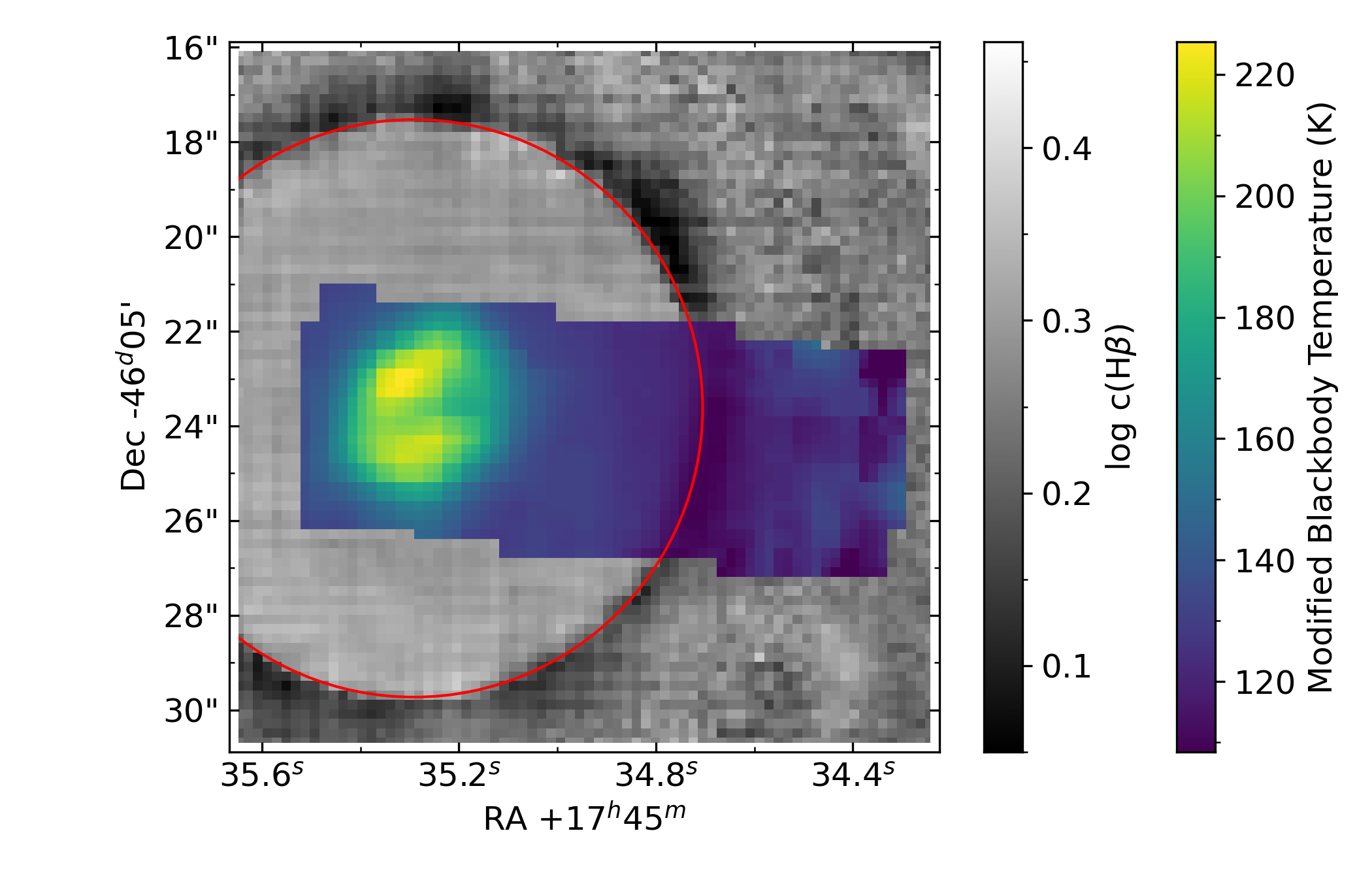}
    \caption{Spatial distribution of the temperature of the blackbody component in the continuum fit, overlaid on the visual extinction (from \citealt{Walsh:MUSE}). The outer edge of the C$_{60}$ ring is illustrated by the red circle.}
    \label{fig: dust temperature vs extinction}
\end{figure}

\cite{Walsh:MUSE} present the morphology of Tc~1 with optical MUSE observations. They find an interesting structure in the extinction: an annulus of extinction lower than the interstellar value, just outside of the ring of C$_{60}$ emission. The blackbody temperature of the continuum fit (Figure \ref{fig: continuum maps}) shows a structure that overlaps with this annulus: the temperature reaches a minimum in this annulus and rises again beyond it. Figure \ref{fig: dust temperature vs extinction} shows the extinction map from \cite{Walsh:MUSE}, with the map of the blackbody temperature overlaid. Obviously, the extinction to any part of the nebula cannot physically be less than the interstellar extinction. \cite{Walsh:MUSE} propose that the dust properties change at the position of the ``negative extinction'' ring. If the particles were larger, then $R = A_V / E(B-V)$ is larger and the extinction could be the same as the halo. 
Regardless of what causes this structure in the optical extinction and the IR continuum, it is no coincidence that it traces the outer edge of the \csixty ring; it must be related to the fullerene formation or excitation.

On the central star, \cite{Walsh:MUSE} find that the extinction is much higher than in the rest of the nebula. We investigate whether this could be due to the presence of circumstellar dust --- e.g. from a circumstellar disk. We define a circular aperture on the central star, with a diameter of 0.7$''$, which is approximately the area where the PSF of the star dominates the surface brightness. Around this circular aperture, we define an annulus-shaped aperture (with an inner diameter of 0.77$''$ and outer diameter of 1.24$''$). The surface brightness in the circular aperture is a combination of the radiation originating from the central star itself, and foreground and background radiation from material in the nebula along the same line of sight. Subtracting the average surface brightness in the annulus from that in the circular aperture leaves the surface brightness originating from the star itself --- assuming the average surface brightness in the annulus is a good approximation of the foreground and background emission. With Tc~1 at a distance of $\sim$3.53 kpc, the diameter of the circular aperture of 0.7$''$ corresponds to a distance of $\sim$2450~AU. Any radiation from circumstellar material within $\sim$1000~AU of the central star would therefore not be spatially resolved and be contained in the PSF of the central star.

\begin{figure}[t]
    \centering
    \includegraphics[width=\linewidth, clip, trim=0.5cm 0.5cm 0.5cm 1cm]{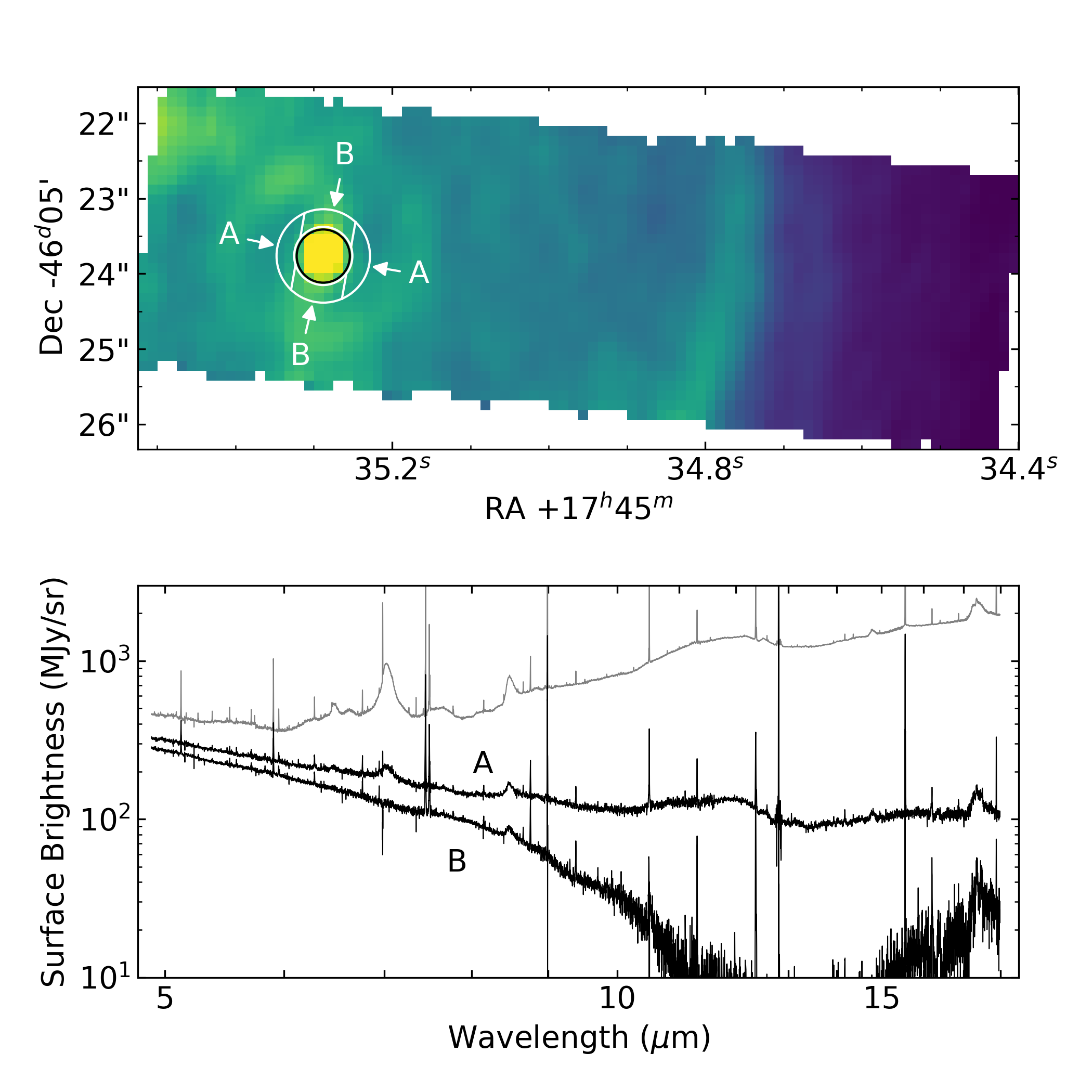}
    \caption{\textit{Top panel}: Average surface brightness between 5 \um and 7.5 \um. The circular aperture on the central star is shown in black. The annulus-shaped aperture around the central star is shown in white; this annulus is divided into 4 areas: the east and west areas make up part A of the annulus, the north and south areas make up part B. \textit{Bottom panel}: Surface brightness on the central star (gray), with the surface brightness of the surrounding annulus --- part A and B respectively --- subtracted (black).}
    \label{fig: star vs annulus}
\end{figure}

As can be seen in Figure \ref{fig: star vs annulus} (top panel), the surface brightness in the annulus is not quite uniform; there is a bar-shaped structure connecting the horseshoe\footnote{The ring around the central star at a distance of $\sim$1.2$''$ corresponds to the horseshoe seen in the dust features. At these shorter wavelengths, it appears as a full ring (see \overviewpaper). It is called the horseshoe in \overviewpaper as well; to stay consistent throughout this paper, we use that name here, despite it showing a full ring.} to the central star from the north and south. We define a rectangular aperture, centered on the star, with a width of 0.69$''$, at a position angle of 10$^\circ$ west of north, and divide the annulus into two parts: the part of the annulus outside of this rectangle (i.e. the west and east side) is part A and the intersection of this annulus with the rectangle (i.e. the north and south side) is part B.

Figure \ref{fig: star vs annulus} (bottom panel) shows the average surface brightness in the circular aperture, with the average surface brightness in the annulus --- part A and B respectively --- subtracted. Using part A of the annulus, that difference follows a power law down to $\sim$10 \um. Part B of the annulus shows effectively the same surface brightness as the circular aperture past $\sim$10 \um; the difference follows a power law down to $\sim$8 \um, and then drops to near-zero past $\sim$10 \um. Also of note is that the surface brightness in part A of the annulus shows small differences in the fullerene emission compared to the circular aperture, whereas part B does not; in the difference using part B, there is only a small trace of the \csixty 8.5 \um feature left, but no trace of any other fullerene feature can be seen. Regardless, using either parts of the annulus, no warm dust can be seen close to the star; up to 8$\sim$10 \um the difference shows nothing but a power law.

\subsection{Dust temperature and Ly$\alpha$ heating}

The dust temperature estimated from the continuum shows a very uniform temperature between the horseshoe and the \csixty ring (see Figures \ref{fig: continuum fits} and \ref{fig: continuum maps}). This indicates the heating of the dust grains might not be dominated by direct stellar radiation. If stellar radiation were the only source of heating for the dust grains, the dust temperature would be expected to roughly follow a $r^{-1/2}$ relation, because the stellar radiation is attenuated by $r^{-2}$ and the thermal emission scales with $T^4$. This clearly does not agree with the observed temperature. Because the dust is in an ionized environment, trapped Ly$\alpha$ photons could be an important source of heating. 

The dust heating by direct stellar radiation is described by
\begin{equation*}
    \Gamma_\star = \pi a^2 \int_0^\infty F_\lambda Q_{\text{abs}} d\lambda
\end{equation*}

\noindent{}with $a$ the grain size, $F_\lambda$ the incoming flux at the location of the dust grain and $Q_{\text{abs}}$ the absorption efficiency of the dust grain. $F_\lambda$ can be calculated from the stellar luminosity $L_\star(\lambda)$:
\begin{equation*}
    F_\lambda = \int_0^{\tau_0(\lambda)} \frac{L_\star(\lambda)}{4\pi r^2} e^{-\tau(\lambda)}d\tau
\end{equation*}

\begin{figure}[t]
    \centering
    \includegraphics[width=\linewidth, clip, trim=0.6cm 0.5cm 0.4cm 0.2cm]{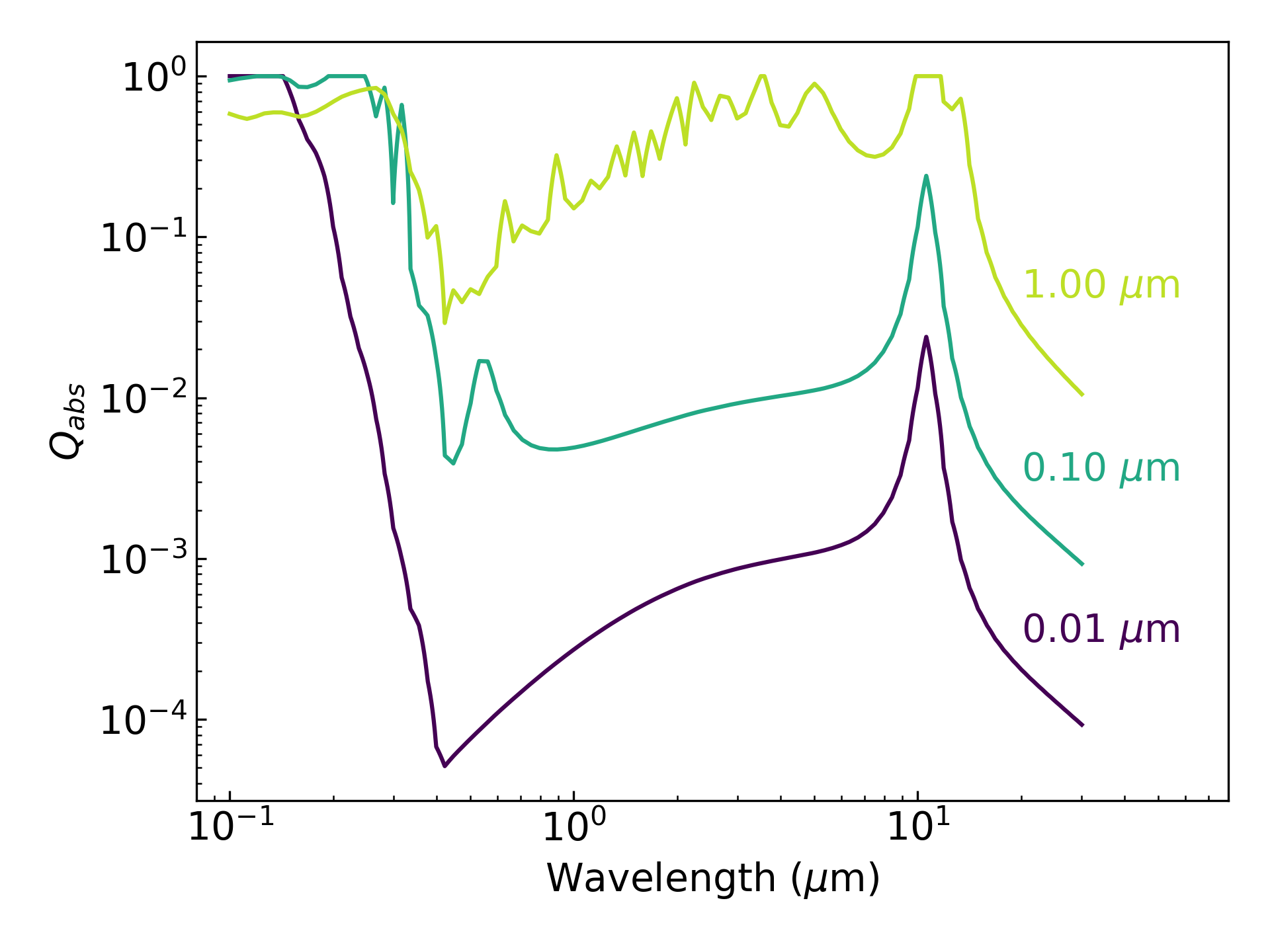}
    \caption{Absorption efficiency of SiC grains for three different sizes \citep{1984ApJ...285...89D}.}
    \label{fig: SiC opacity}
\end{figure}

\noindent{}with $\tau_0(\lambda)$ the optical depth to the star. $Q_{\text{abs}}$ is dependent on wavelength and the size and composition of the grain. Considering SiC grains, the optical properties have been calculated by \cite{1984ApJ...285...89D}; Figure \ref{fig: SiC opacity} shows the absorption efficiency $Q_{\text{abs}}(\lambda)$ of SiC for three different sizes of grains. The opacity is approximately 1 in the UV and drops by 2--4 orders of magnitude past 0.2--0.3 \um. The opacity at longer wavelengths is mostly inconsequential, as the vast majority of energy from the star is emitted in the UV/optical. 

The dust heating by Ly$\alpha$ photons is \citep{2016ApJ...830..118S}
\begin{equation*}
    \Gamma_{\text{Ly}\alpha} = n_e n_H f \beta_B h\nu_\alpha / n_d
\end{equation*}

\noindent{}with $n_e$ the electron density, $n_H$ the hydrogen density, $f=0.7$ the fraction of of recombination to levels higher than two that lead to emission of a Ly$\alpha$ photon, $\beta_B$ the recombination coefficient of hydrogen (case B), $h\nu_\alpha$~=~10.2 eV the energy of a Ly$\alpha$ photon and $n_d$ the dust density. 

These two heating mechanisms are in equilibrium with the cooling, which we consider to be solely due to thermal radiation:
\begin{equation*}
    \Gamma_{\text{em}} = 4 \pi a^2 Q_{\text{em}} \sigma {T_d}^4
\end{equation*}

\noindent{}with $\sigma$ the Stefan-Boltzmann constant and $T_d$ the dust temperature. $Q_{\text{em}}$ is the emission efficiency, integrated over a blackbody spectrum with temperature $T_d$. For grains smaller than 0.3 \um and 65~K~$<$~$T_d$~$<$~250~K, $Q_{\text{em}}$ can be approximated as \citep{tielens_ism_2005}
\begin{equation*}
    Q_{\text{em}} \approx 6\times 10^{-2} \left( \frac{a}{1~\upmu\text{m}} \right)
\end{equation*}

\noindent{}Equating the heating sources with the cooling, the dust temperature is
\begin{equation}\label{eq: dust temperature}
    {T_d}^4 = \frac{1}{4 Q_{\text{em}} \sigma} \left[ \frac{n_e n_H f \beta_B h\nu_\alpha}{\pi a^2 n_d} + \int_0^\infty F_\lambda Q_{\text{abs}} d\lambda \right]
\end{equation}

We use a spectral energy distribution (SED) from a Kurucz model \citep{1979ApJS...40....1K} at $T_{\text{eff}} = 31000$ K and $\log g = 3.5$ \citep{2019MNRAS.490.2475A, Walsh:MUSE} to model the radiation of the central star. We scale the SED to match the luminosity of the central star. \cite{Walsh:MUSE} report a luminosity of $L_\star=8150 L_\odot$. We find that this is consistent with the adopted stellar and environmental parameters. We find no evidence of an ionization front in Tc~1 (\overviewpaper), thus the central star must be bright enough to fully ionize the nebula. With the adopted stellar parameters we find a Str\"omgren radius of 0.09~pc, or 5.3$''$---roughly equal to the \csixty ring\footnote{H$\alpha$ emission is seen throughout the entire 55$''$ diameter halo of Tc~1, but the density is much lower outside the \csixty ring. We thus consider that very few ionizing photons are required to ionize the halo compared to the core of Tc~1.}.

The extinction has been determined by \cite{Walsh:MUSE}; they find a uniform extinction within the core of $c(\text{H}\beta) = 0.306 \pm 0.023$. This includes interstellar extinction, which they determined to be $c(\text{H}\beta) = 0.23$, assuming a Seaton reddening law with $R_V=3.1$. The internal extinction in the core is thus $c(\text{H}\beta) = 0.076$. The reddening law of the internal extinction is uncertain. We adopt the same Seaton reddening law for internal extinction, but since the dust composition in Tc~1 is different from that in the diffuse ISM, this may not be a good approximation. The far-UV rise may vary significantly depending on the dust composition. Under this assumption, the internal extinction of $c(\text{H}\beta) = 0.076$ correspond to $A_V = 0.16$.

To determine the internal extinction up to a distance $r$ from the central star, we estimate that the typical distance a photon travels inside of the core is equal to the radius of the core (corresponding to about 6$''$). We thus estimate that the extinction per distance traveled in the core is about $A_V/r = 0.027$ per arcsecond, where we make the approximation that the distance to the central star is equal to the projected distance in the plane of the sky.

\begin{figure}[t]
    \centering
    \includegraphics[width=\linewidth]{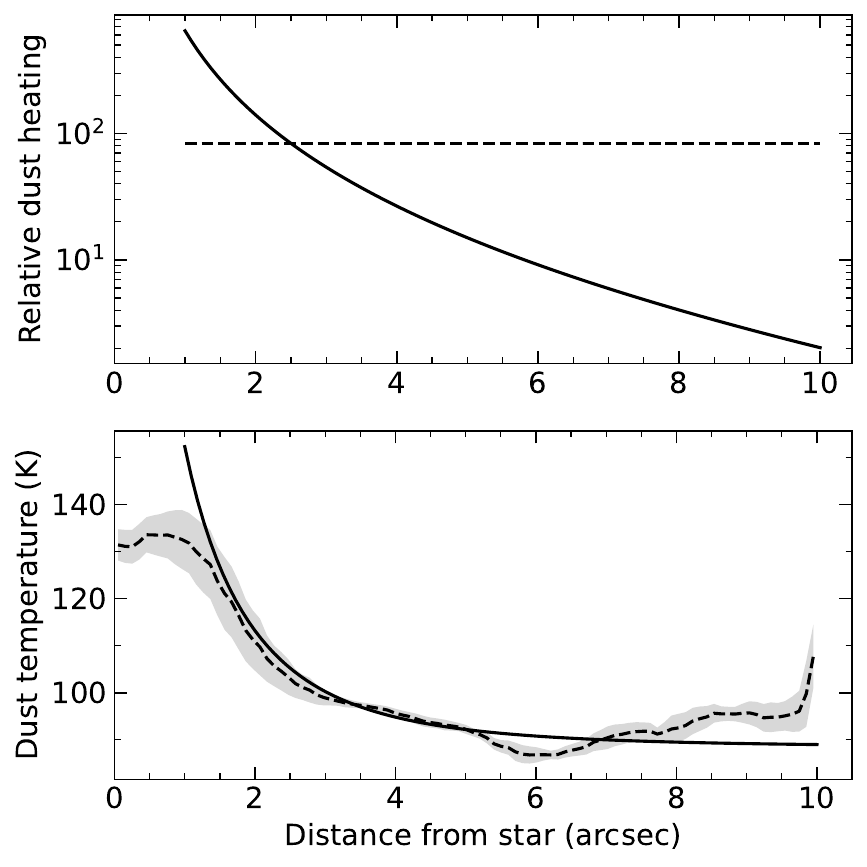}
    \caption{\textit{Top panel:} Relative contribution of heating from direct stellar radiation (solid) and from Ly$\alpha$ (dashed). \textit{Bottom panel:} Predicted dust temperature (solid), compared to the dust temperature inferred from the continuum fits (dashed). The azimuthal standard deviation of the inferred dust temperature is shown in gray.}
    \label{fig: dust temperature model}
\end{figure}

We adopt $a$~=~0.1~\um and find that for a dust abundance of $n_H/n_d$~=~$2.5 \times 10^{-13}$, the predicted dust temperature matches the temperature inferred from the continuum fits (Figure \ref{fig: dust temperature model}). We find that up to $\sim$2$''$ from the central star, the heating is dominated by direct stellar radiation, which may explain the higher temperature seen in the horseshoe. The horseshoe structure itself, however, is not explained by this simple model. We conclude that the relatively constant dust temperature past the horseshoe can be explained by Ly$\alpha$ heating.

\section{Conclusions}\label{sec: conclusion}
We study the infrared emission from dust in the planetary nebula Tc~1, using MIRI/MRS observations, to characterize the environment in which fullerenes are formed. The dust emission in all fullerene-rich PNe shows similar continuum emission, and 3 broad features: the 6--9~\um plateau, the 9--13~\um plateau and the 30~\um feature. From a combined spectral and spatial analysis, we find that the spectral shape of the 6--9~\um plateau is constant throughout the FOV, whereas the 9--13~\um plateau shows a large variation in shape.

We decompose the 9--13 ~\um plateau and find that 3 components are sufficient to reproduce the variations in the shape. We find that two of the components (peaking at 11~\um and 12~\um) strongly resemble SiC \citep{HENNING2001815, 2008A&A...483..661P, 2014ApJ...791...28S}, and thus conclude that one of the contributors to the 9--13~\um plateau is likely SiC. The two different components must be produced by grains that differ in some way (e.g. shape, size or surface composition); the different spatial distribution of these two components therefore hints at dust processing taking place in the nebula. For the third component of the 9--13~\um plateau (at 9.5~\um) we tentatively propose similar materials (e.g. TiC or ZrC) as promising candidates for the carrier species.

We investigate the spatial distribution of the integrated surface brightness of the various features and continuum emission and find that the two plateaus have distinctly different distributions in the nebula, indicating the carriers of these plateaus are different species. The 6--9~\um plateau shows a different distribution from all other dust features and the continuum emission. For this reason, in addition to the spectral shape of the 6--9~\um plateau being constant throughout the FOV, we conclude that the carrier of the 6--9~\um plateau is likely stochastically heated material. We propose carbonaceous molecules --- whether in the form of large PAHs, large fullerenes or species with similar (sub)structures --- as promising carrier candidates.

From a global continuum fit, we estimate the dust temperature. We compare this to a simple model and find that past $\sim$2$''$ from the central star, the dust heating is dominated by trapped Ly$\alpha$ photons, leading to a relatively constant dust temperature throughout the rest of the field of view.

\begin{acknowledgments}
This work is based on observations made with the NASA/ESA/CSA James Webb Space Telescope. All of the data presented in this article were obtained from the Mikulski Archive for Space Telescopes (MAST) at the Space Telescope Science Institute. The data of this specific observing program can be accessed via \dataset[doi:10.17909/9p7c-fn82]{https://archive.stsci.edu/doi/resolve/resolve.html?doi=10.17909/9p7c-fn82}.
This article is based upon work from COST Action CA21126 - Carbon molecular nanostructures in space (NanoSpace), supported by COST (European Cooperation in Science and Technology).
M.~G., J.~C., S.~V.S., E.~P., C.~B., and D.~V.D.P. acknowledge support from the University of Western Ontario, the Canadian Space Agency (CSA) [24JWGO3A13], and the Natural Sciences and Engineering Research Council of Canada. V.~J.~E thanks Chapman University for support. D.~A.~G.~H. and A.~M. acknowledge the support from the State Research Agency (AEI) of the Ministry of Science, Innovation and Universities (MICIU) of the Government of Spain, and the European Regional Development Fund (ERDF), under grant PID2023-147325NB-I00/AEI/10.13039/501100011033. Astrochemistry in The Netherlands is supported by the NWO Dutch Astrochemistry Network (grant no. ASTRO.JWST.001). R.~S.’s contribution to the research described here was carried out at the Jet Propulsion Laboratory, California Institute of Technology, under a contract with NASA (80NM0018D0004). A.~G.~G.~M.~T., G.~C.~S., H.~L.~D., K.~E.~K., R.~S., and N.~C.~S.
received support from award JWST-GO-04706 from the STScI under NASA
contract NAS5-03127. K.J. acknowledges the support of the Swedish National Space Agency. 
Portions of the text in this paper were drafted and edited with the assistance of
Claude, a large language model developed by Anthropic
(\url{https://claude.ai}). The
tool was used under the direction of the authors, who verified all content and take
full responsibility for the analysis, results and conclusions presented here.
\end{acknowledgments}

\software{
    JWST Pipeline \citep{bushouse_2025_14597407}, 
    \textsc{astropy} \citep{2013A&A...558A..33A, 2018AJ....156..123A},  
    \textsc{scipy} \citep{SciPy2020}, 
    \textsc{lmfit} \citep{newville_2025_15014437}
}

\bibliography{refs}{}
\bibliographystyle{aasjournalv7}

\appendix
\twocolumngrid

\section{Additional Tables}\label{sec: additional figures and tables}
Table \ref{tab: aperture centers} shows the centers of the square apertures used in section \ref{sec:dustfeatures}.

\begin{table}[h]
    \centering
    \caption{The center locations of the apertures shown in figure \ref{fig:apertures preview}. Each aperture is 0.7'' in size.}
    \begin{tabular}{cc}
        \hline
        \hline
        Aperture & Center \\
        \hline
        1 & $(17^{\text{h}}45^{\text{m}}35.296^{\text{s}}, -46^{\text{d}}05'24.202'')$ \\
        2 & $(17^{\text{h}}45^{\text{m}}35.228^{\text{s}}, -46^{\text{d}}05'24.202'')$ \\
        3 & $(17^{\text{h}}45^{\text{m}}35.161^{\text{s}}, -46^{\text{d}}05'24.202'')$ \\
        4 & $(17^{\text{h}}45^{\text{m}}35.094^{\text{s}}, -46^{\text{d}}05'24.202'')$ \\
        5 & $(17^{\text{h}}45^{\text{m}}35.026^{\text{s}}, -46^{\text{d}}05'24.202'')$ \\
        6 & $(17^{\text{h}}45^{\text{m}}34.959^{\text{s}}, -46^{\text{d}}05'24.202'')$ \\
        7 & $(17^{\text{h}}45^{\text{m}}34.892^{\text{s}}, -46^{\text{d}}05'24.202'')$ \\
        8 & $(17^{\text{h}}45^{\text{m}}34.825^{\text{s}}, -46^{\text{d}}05'24.202'')$ \\
        9 & $(17^{\text{h}}45^{\text{m}}34.757^{\text{s}}, -46^{\text{d}}05'24.202'')$ \\
        10 & $(17^{\text{h}}45^{\text{m}}34.690^{\text{s}}, -46^{\text{d}}05'24.202'')$ \\
        11 & $(17^{\text{h}}45^{\text{m}}34.623^{\text{s}}, -46^{\text{d}}05'24.202'')$ \\
        \hline
    \end{tabular}
    \label{tab: aperture centers}
\end{table}

\section{The molecular features and their influence on the plateaus}\label{sec: appendix molecular features}

\subsection{List of the molecular features}\label{sec: list of molecular features}

Table \ref{tab: molecular features} provides a list of molecular features that are subtracted from the data in the analysis presented in Section \ref{sec:dustfeatures}. Most features are fitted with Gaussian profiles. The fit of the C$_{60}$ 7.0 \um feature is taken from \morganpaper, who use two Lorentzian profiles. The 11.2 \um feature is fitted with a skewed Gaussian with the following expression:
$$ f = A \exp \left( \frac{(\lambda - \lambda_0)^2}{2 \Sigma^2} \right) $$
$$ \Sigma =
\begin{cases} 
  (1 - \tanh(\gamma)) \sigma & \lambda \leq \lambda_0 \\
  (1 + \tanh(\gamma)) \sigma & \lambda > \lambda_0 \\
\end{cases}
$$
\noindent{}with peak position $\lambda_0$, amplitude $A$, standard deviation $\sigma$ and skewness $\gamma$. The skewness is parametrized such that the total integral of this function is $\sqrt{2 \pi} A \sigma$. The skewness $\gamma$ is defined on $[-\infty, \infty]$; $\gamma = 0$ corresponds to a symmetric Gaussian with standard deviation $\sigma$, $\gamma < 0$ produces a wider left wing and $\gamma > 0$ produces a wider right wing. For the fit of the 11.2~\um feature $\gamma = 0.4$ was used.

\begin{table}[t]
    \centering
    \caption{List of molecular features subtracted from the data in the analysis presented in Section \ref{sec:dustfeatures}. The second column indicates the shape of the profile that was used for fitting: Gaussian (``G''), skewed Gaussian (``SG'') or Lorentzian (``L'').}
    \begin{tabular}{cccc}
        \hline
        \hline
        Central & Shape & FWHM & Identification \\
        wavelength &  & (\um) &  \\
        (\um) &  &  &  \\
        \hline
        5.05 & G & 0.077 & \csixty \\
        6.23 & G & 0.10 & PAH \\
        6.40 & G & 0.041 & Unidentified \\
        6.49 & G & 0.065 & \csixty \\
        6.64 & G & 0.13 & Unidentified \\
        7.01 & L & 0.095 & \csixty \\
        7.05 & L & 0.186 & \csixty \\
        7.20 & G & 0.10 & Unidentified \\
        7.63 & G & 0.25 & Unidentified \\
        8.47 & G & 0.12 & \csixty \\
        8.83 & G & 0.071 & \cseventy \\
        9.62 & G & 0.082 & Unidentified \\
        9.97 & G & 0.35 & Unidentified \\
        10.47 & G & 0.16 & Unidentified \\
        11.23 & SG & 0.35 & PAH \\
        12.52 & G & 0.082 & \cseventy \\
        12.60 & G & 0.16 & Unidentified \\
        14.05 & G & 0.22 & Unidentified \\
        14.39 & G & 0.28 & Unidentified \\
        14.79 & G & 0.12 & \cseventy \\
        \hline
    \end{tabular}
    \label{tab: molecular features}
\end{table}

\subsection{Influence of the C$_{60}$ 7.0 $\mu$m band on the 6-9 $\mu$m plateau}

It is important to consider how well the local continuum of the C$_{60}$ 7.0 \um band is constrained, because it directly affects the uncertainty on the 6--9 \um plateau. The local continuum of the C$_{60}$ 7.0 \um band consists of the combined contribution of the 6--9 \um plateau and the dust continuum emission. Since the C$_{60}$ 7.0 \um band is by far the strongest feature in the 6--9 \um range, this fit will be the dominant source of systematic uncertainty caused by the subtraction of the molecular features. Additionally, the profile of the 7.0 \um band is Lorentzian in shape, which has very broad wings. With the other molecular features on top of the wings and the unknown shape of the 6--9 \um plateau potentially varying on scales similar to the width of the 7.0 \um band, the only constraint on the local continuum is the precise shape of the 7.0 \um band. The influence on the 6--9 plateau of the uncertainty of the local continuum under 7.0 \um band is therefore second only to the uncertainty of the dust continuum emission.

While it is infeasible to provide any realistic estimates of systematic uncertainties on the fit of the 7.0 \um band, it is possible to estimate the statistical uncertainty. In \morganpaper we fit the C$_{60}$ 7.0 \um band with 5 components: two Lorentzian profiles for the C$_{60}$ band, two Gaussian profiles for the 6.9 and 7.2 features and a linear local continuum. In \morganpaper we fix the width and location of the various components to produce spatial maps, which helps greatly in converging reliably to a good fit in every pixel, but in reality none of these parameters are known. We thus allow all parameters to vary in the following analysis.

\begin{figure}[t]
    \centering
    \includegraphics[width=\linewidth, clip, trim=0.5cm 0cm 0cm 0cm]{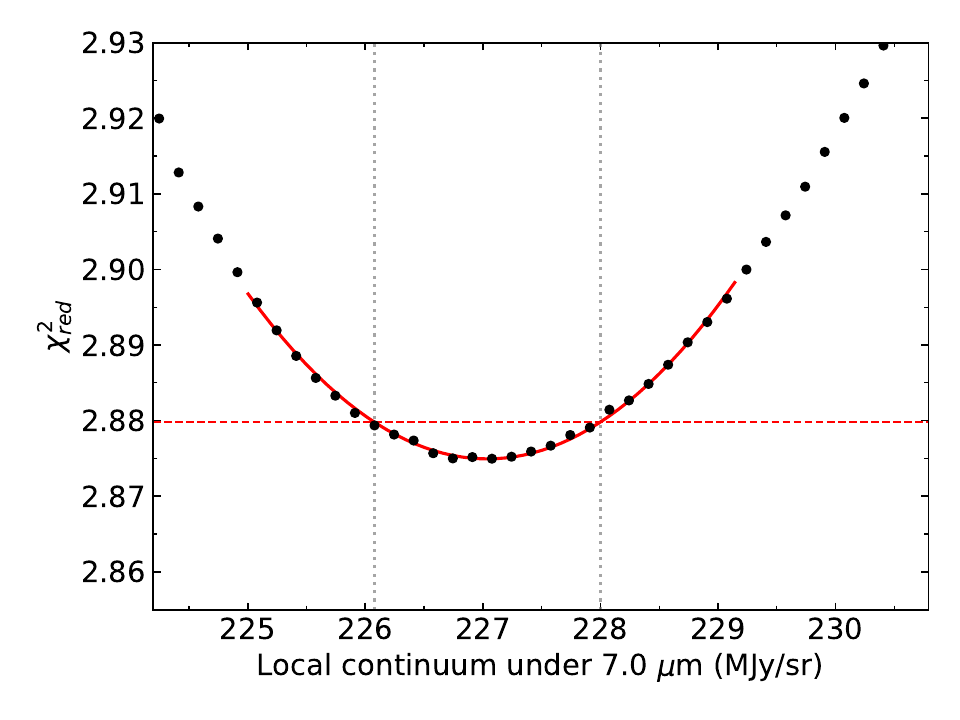}
    \caption{Reduced $\chi^2$ landscape of the local continuum of the C$_{60}$ 7.0 \um band at 7.03 \um (black dots), with a quadratic fit (solid red). The $\chi^2$ value corresponding to a 1-$\sigma$ confidence interval is indicated with a red dashed line. The corresponding confidence interval of the local continuum is indicated with dotted gray lines.}
    \label{fig:local continuum at 7 um}
\end{figure}

To estimate the statistic uncertainty of the local continuum at 7.0~\um, we fit the surface brightness in aperture 4\footnote{Aperture 4 was chosen as it shows the ``typical'' case for the strength of fullerene-, plateau- and continuum emission in Tc~1.}, between 6.77~\um and 7.33~\um, while fixing the local continuum at 7.03~\um to a value $b$. Performing this fit for a range in values of $b$ produces a one-dimensional $\chi^2$ landscape, as shown in Figure \ref{fig:local continuum at 7 um}. The $\chi^2$ landscape is fitted with a quadratic polynomial close to the minimum, and from the intersections with the $\chi^2$ value corresponding to the 1-$\sigma$ confidence interval the uncertainty on the local continuum is determined. This leads to a statistical uncertainty of $\sim$1~MJy/sr. The statistical uncertainty on the local continuum at 7.03~\um is thus very small, and it is safe to assume the uncertainty on the surface brightness of the 6--9~\um plateau is dominated by the systematic uncertainties of the global continuum and the fit of the C$_{60}$ 7.0~\um band.

\end{document}